\UseRawInputEncoding
\documentclass[pra,twocolumn,groupedaddress,showpacs,floatfix]{revtex4}

\usepackage{graphics}
\usepackage{graphicx}
\usepackage{bm}
\usepackage{amsmath}
\usepackage{amsfonts} 
\usepackage{amssymb}
\usepackage{latexsym}
\usepackage{color}
\usepackage{mathrsfs}
\usepackage{braket}
\usepackage{enumerate}
\usepackage[usenames,dvipsnames]{xcolor}
\usepackage[colorlinks=true,citecolor=blue,linkcolor=blue,urlcolor=blue,backref=true]{hyperref}
\usepackage{float}
\usepackage{mathtools}
\usepackage[raggedright,tight]{subfigure}  
\usepackage[inline]{enumitem}
\usepackage{tikz}
\usetikzlibrary{matrix}
\usepackage{natbib}

\begin{document}

\title{Ultrastrong time-dependent light-matter interactions are gauge-relative}
\author{Adam Stokes}\email{adamstokes8@gmail.com}
\affiliation{Department of Physics and Astronomy, University of Manchester, Oxford Road, Manchester M13 9PL, United Kingdom}
\author{Ahsan Nazir}\email{ahsan.nazir@manchester.ac.uk}
\affiliation{Department of Physics and Astronomy, University of Manchester, Oxford Road, Manchester M13 9PL, United Kingdom}

\date{\today}

\begin{abstract}
Time-dependent light-matter interactions are a widespread means by which to describe controllable experimental operations. They can be viewed as an approximation in which a third system - the control system - is treated as external within the Hamiltonian. We demonstrate that this results in non-equivalence between gauges. We provide a physical example in which each different non-equivalent model coincides with a gauge-invariant description applied in a different experimental situation. The qualitative final-time predictions obtained from these models, including entanglement and photon number, depend on the gauge within which the time-dependent coupling assumption is made. This occurs whenever the interaction switching is sufficiently strong and non-adiabatic even if the coupling vanishes at the preparation and measurement stages of the protocol, at which times the subsystems are unique and experimentally addressable.
\end{abstract}


\maketitle

\section{Introduction}

Exploiting controlled light-matter coupling is important for quantum computation \cite{nielsen_universal_2002,haroche_exploring_2006,zhang_conditions_2005,ladd_quantum_2010,romero_ultrafast_2012}, quantum communication \cite{gisin_quantum_2002}, quantum metrology \cite{giovannetti_advances_2011,bayer_terahertz_2017}, and quantum simulation \cite{feynman_simulating_1982,zeier_gate_2004}. In the search for scalable platforms operating at room-temperature, strong light-matter coupling has become of major interest through solid-state systems, such as semiconductor quantum wells \cite{weisbuch_observation_1992,liberato_quantum_2007} and dots \cite{reithmaier_strong_2004,hennessy_quantum_2007}, through two-dimensional \cite{lundt_room-temperature_2016} and organic \cite{daskalakis_spatial_2015} materials, and through superconducting circuits \cite{schuster_resolving_2007,gunter_sub-cycle_2009,niemczyk_circuit_2010,peropadre_switchable_2010,yoshihara_superconducting_2017,romero_ultrafast_2012,armata_harvesting_2017}.


Here we consider the implications of the widespread and important practice of modelling controllable light-matter interactions by assuming that the coupling parameter $\eta$ of the model Hamiltonian, $H(\eta)$, depends on an external control parameter \cite{leroux_enhancing_2018,keyl_controlling_2014,forn-diaz_ultrastrong_2019,romero_ultrafast_2012,wang_fast_2004,de_liberato_extracavity_2009,xiao_fast_2019,garziano_switching_2013,kyaw_scalable_2015,zanardi_quantum_2004,filipp_two-qubit_2009,peropadre_switchable_2010,deppe_two-photon_2008,hofheinz_generation_2008,hofheinz_synthesizing_2009,mariantoni_two-resonator_2008,johansson_dynamical_2009,wang_coupling_2009,kerman_high-fidelity_2008}. This means that $\eta$ varies in time; $\eta\to \eta(t)$. For example, time-dependent couplings in cavity QED can be used to realise a universal set of gates for quantum computation \cite{keyl_controlling_2014} and ultrastrong ultrafast couplings are proposed to realise high-fidelity gates using superconducting circuits \cite{romero_ultrafast_2012}. Such models may also result from the rotation of a model in which a subsystem is classically driven (e.g. \cite{forn-diaz_ultrastrong_2019}). Time-dependent Hamiltonian components are of intrinsic importance in thermodynamics, where they are used to define work as a component of energy. They are also important in optimal control theory, through models such as the extended Rabi model \cite{shore_jaynes-cummings_1993,ruggenthaler_quantum-electrodynamical_2014,pellegrini_optimized_2015,flick_atoms_2017}. The assumption of preparing or measuring an eigenstate of the non-interacting Hamiltonian even when interactions are present is also common, and this is equivalent to assuming an instantaneous switching of the interaction (e.g. \cite{power_coulomb_1959,milonni_natural_1989,power_time_1999,stokes_gauge_2013}). 

The assumption of a time-dependent coupling, or of preparing and measuring particular non-interacting states, is highly non-trivial because light and matter quantum subsystems are defined differently by different gauges \cite{stokes_implications_2020}. Within traditional weak-coupling regimes focus has predominantly been placed on the Coulomb and multipolar gauges \cite{drummond_unifying_1987,stokes_extending_2012,stokes_gauge_2013,stokes_master_2018,stokes_gauge_2019}. The Pauli-Fierz representation, which attempts to isolate the component of the electromagnetic field tied to the material system, has also been used to calculate radiative corrections \cite{cohen-tannoudji_atom-photon_2010}. These previous studies have focussed specifically on establishing gauge-invariance of the $S$-matrix \cite{fried_vector_1973,bassani_choice_1977,forney_choice_1977,kobe_question_1978,cohen-tannoudji_photons_1997,craig_molecular_1998,woolley_r._g._charged_1999,woolley_gauge_2000}, or else have considered the natural lineshape problem of spontaneous emission in weak-coupling and Markovian regimes \cite{Lamb_fine_1952,low_natural_1952,power_coulomb_1959,milonni_natural_1989,power_time_1999,stokes_extending_2012,stokes_gauge_2013}. It is now well-known that QED $S$-matrix elements calculated using perturbation theory are independent of the subsystem division at every order \cite{cohen-tannoudji_photons_1997,woolley_r._g._charged_1999,woolley_gauge_2000}. This result is physically limited however, because it is a direct consequence of the adiabatic switching condition definitive of the $S$-matrix \cite{cohen-tannoudji_photons_1997}.

In scattering theory virtual processes are allowed only as intermediates within a ``real process". On the other hand, beyond  scattering theory, virtual effects are especially important in ultrastrong-coupling regimes and when dealing with ultrafast interactions. Moreover, ultrastrong light-matter coupling is now a major field of study for both fundamental and applied physics \cite{forn-diaz_ultrastrong_2019,kockum_ultrastrong_2019}. Likewise, increasingly fast interactions are becoming more and more prevalent, and in particular, may be advantageous in mitigating detrimental environmental affects occurring over the course of a controlled process. Sub-cycle interaction switching was in fact achieved some time ago \cite{gunter_sub-cycle_2009} and more recently, sub-optical-cycle dynamics have been achieved within the ultrastrong coupling regime by exploiting vacuum fluctuations rather than coherent driving \cite{halbhuber_non-adiabatic_2020}. 

Here we consider controllable light-matter interactions, but we avoid the restrictive scattering-theoretic assumption of adiabatic switching over infinite times, as is required for modelling any platform that involves sufficiently fast and strong interaction switching. We also allow the gauge to be arbitrary. Since each gauge for $H(\eta)$ provides a different physical definition of the interacting  ``light" and ``matter" quantum subsystems, the promotion $\eta \to \eta(t)$ constitutes a different physical assumption when applied in each different gauge. Moreover, beyond scattering theory, different gauges generally treat virtual processes very differently within sufficiently fast and strong interaction regimes. We show that as a result of this, in such regimes the final-time predictions obtained from $H(\eta(t))$ are significantly different in different gauges. This occurs even when the quantum subsystems are unique at the preparation and measurement stages at which times the interaction vanishes. 

We show using the example of an atom moving through a cavity that by including from the outset an explicit description of the degrees of freedom that mediate the interaction, i.e., by explicitly including the control system at the Lagrangian level, one can obtain a more complete and gauge-invariant description ${\tilde H}(t)$. We demonstrate that each different model $H(\eta(t))$, each of which belongs to a different gauge, coincides with ${\tilde H}(t)$ applied to a different microscopic arrangement of the overall system. Thus, when using the assumption of a time-dependent-coupling, $\eta\to \eta(t)$, each gauge models a specific microscopic arrangement. These findings place significant restrictions on the validity of using this common method when describing strong and fast interactions, because determining which specific experimental arrangements a particular model $H(\eta(t))$ describes, requires a more complete theory, such as ${\tilde H}(t)$, which may be unavailable or intractable except in the simplest cases.

The paper is organised into five sections. In Sec.~\ref{cont} we provide theoretical background introducing time-dependent interactions. In Sec.~\ref{hd} we provide a simple toy model that transparently demonstrates the implications of using a time-dependent coupling parameter. In Sec.~\ref{cavity} we consider a more realistic atom-cavity system which facilitates a comparison between the time-dependent coupling method and more complete descriptions of the controlled light-matter interaction. In Sec.~\ref{general} we consider the time-dependent coupling method when describing fast and strong interactions. Finally we briefly summarise our findings in Sec.~\ref{conc}.

\section{Controllable electromagnetic interactions}\label{cont}

\subsection{Definition of interaction and external control approximation}

Maxwell's equations can be derived from the standard QED Lagrangian
\begin{align}\label{Lag}
{\cal L} = L_{\rm m}- \int d^3 x \left[j_\mu A^\mu + {1\over 4}F_{\mu\nu}F^{\mu\nu}\right] 
\end{align}
where $L_{\rm m}$ and $j$ are the free Lagrangian and the four-current of an arbitrary material system such that $dj=0$, while $F=dA$ and $A$ are the electromagnetic tensor and four-potential for an arbitrary electromagnetic system. All fields are assumed to vanish sufficiently rapidly at the boundaries of the integration domain. Under a gauge transformation $A \to A -d\chi$, where $\chi$ is arbitrary, the Lagrangian is transformed to one that differs by a total time-derivative, and which is therefore equivalent to ${\cal L}$.

In applications we often wish to control the interaction between light and matter systems, such as atoms within a cavity. This control can only occur via a third system, such as a laser, which we call the control subsystem. Often, the explicit inclusion of this control subsystem via a fully quantum treatment is cumbersome or even intractable. In this case a simpler alternative must be sought. The simplest approach is to promote the light-matter coupling parameter to a time-dependent function, which then constitutes an implicit model for the control subsystem. One could instead pursue an explicit description of the control subsystem as classical and external, i.e., as possessing pre-prescribed dynamics. This external control approximation could be implemented at either the Lagrangian level or the Hamiltonian level, but it is not clear that these two approaches will be equivalent.


Even if one confines oneself to attempting to treat the control subsystem as external, in many situations a tractable model may still be unavailable. We therefore start here by considering the simpler approach of using a time-dependent light-matter coupling parameter, which is a widespread approach within the literature. Our aim is to understand the implications of this approach when dealing with very fast and strong interactions. In due course, we will see within specific examples how, if at all, this description differs from explicitly modelling the control subsystem as external. 

Our first task is to define what is meant by an interaction. The definition must be such that when the interaction vanishes the theory reduces to two free theories. A natural approach to describing time-dependent interactions would be to modify the interaction Lagrangian density ${\mathscr L}_I=-j_\mu A^\mu$ via the replacement ${\mathscr L}_I \to \mu(t){\mathscr L}_I$, where $\mu(t)$ is a time-dependent coupling function. However, this alone does not imply that the interaction vanishes when $\mu(t)=0$, due to Gauss' law $\nabla \cdot {\bf E}=\rho$, where $\rho=j_0$ and $E_i =F^{0i}$. Instead, a modified current may be defined as $\mu(t)j$. Whenever $\mu(t)=0$ one then recovers two independent and free theories with matter described by $L_{\rm m}$, and the electromagnetic subsystem described by ${\mathscr L}_{\rm TEM} = ({\bf E}_{\rm T}^2-{\bf B}^2)/2$ where ${\bf E}={\bf E}_{\rm T}$ is transverse and ${\bf B}=\nabla\times {\bf A}$ is the magnetic field.

Solving $\nabla \cdot {\bf E}=\mu(t)\rho$ to obtain ${\bf E}_{\rm L}$, and replacing $j_\mu$ with $\mu(t) j_\mu$ in Eq.~(\ref{Lag}) yields the Lagrangian
\begin{align}\label{lag2}
L = &L_{\rm m} +L_{\rm TEM}+{\mu(t)^2 \over 2}\int d^3 x \, \rho \phi_{\rm Coul} \nonumber \\ & -\mu(t) \int d^3 x \left[ \rho A_0 - {\bf J}\cdot {\bf A}\right] 
\end{align}
where $J_i=j^i,~i=1,\,2,\,3$ and
\begin{align}
\phi_{\rm Coul}({\bf x}) = \int d^3 x' {\rho({\bf x}') \over 4\pi |{\bf x}-{\bf x}'|}.
\end{align}
This formulation accommodates an arbitrary time-dependent interaction, arbitrary material and electromagnetic systems, and arbitrary choice of gauge.

\subsection{Non-equivalent Lagrangians}\label{noneqlag}

We now consider a gauge transformation $A_\mu \to A_\mu -\partial_\mu\chi$, which in terms of scalar and vector potentials reads
\begin{align}\label{gauge}
A_0 \to A_0 -\partial_t \chi,\qquad  {\bf A}\to {\bf A}+\nabla \chi.
\end{align}
This transforms the Lagrangian $L$ in Eq. (\ref{lag2}) as
\begin{align}\label{t1}
L \to L + \mu(t){d\over dt} \int d^3x \rho \chi.
\end{align}
The right-hand-side is equivalent to $L$ if and only if ${\dot \mu}=0$.

Since the total electric charge is the conserved Noether-charge resulting from gauge-symmetry, the non-equivalence can be understood as a consequence of the fact that if $\partial_\mu j^\mu=0$ then $\partial_\nu \mu(t)j^\nu= 0$ if and only if ${\dot \mu}=0$. One is naturally led to seek a different modified current $\tilde j$, which includes the external control $\mu(t)$, but also satisfies $\partial_\nu {\tilde j}^\nu=0$. We perform this analysis in Appendix A. The construction of ${\tilde j}$ requires inverting the divergence operator, which introduces an additional, equally significant, gauge arbitrariness into the formalism. Neither $\mu(t)j$ nor ${\tilde j}$ results in a Lagrangian that provides invariant dynamics under a complete gauge transformation. Thus, introducing the single additional assumption that the interaction is controllable, has resulted in non-equivalence between different gauges. 

As we have noted, a more complete approach would include an explicit model for the control subsystem. In Sec.~\ref{tdh} we will consider the example of an atom moving in and out of a cavity, which is simple enough to be amenable to such an approach. In this case the control subsystem is the atom's centre-of-mass motion. We will see that approximating the control subsystem as external at the Hamiltonian level actually produces the same result as assuming a time-dependent coupling parameter. Thus, the non-equivalence of models belonging to different gauges when using the time-dependent coupling method, can be understood as the consequence of an approximation. The situation is analogous to the effect of material energy-level truncation, which also produces non-equivalent models when applied in different gauges \cite{de_bernardis_breakdown_2018,stokes_gauge_2019,di_stefano_resolution_2019,roth_optimal_2019}. However, while material level truncation is often straightforwardly avoidable, avoiding the approximation of a control system as being external may be much more difficult. 

In Sec.~\ref{gih} we will see that if we instead approximate the control subsystem as external at the Lagrangian level, then the subsequent Hamiltonians belonging to different gauges are equivalent. Furthermore, through comparison with this Lagrangian approach it is possible to determine whether or not the simple time-dependent coupling method will be valid when applied within a given gauge. We find that the correct gauge, if any, to employ when using the latter method, depends on the microscopic details of the system. The procedure is analogous to identifying a gauge that provides the most accurate material truncation by comparing the approximate non-equivalent models with a more complete gauge-invariant theory \cite{de_bernardis_breakdown_2018,stokes_gauge_2019,di_stefano_resolution_2019,roth_optimal_2019}.

\subsection{Non-equivalent Hamiltonians}\label{nonham}

To better understand the implications of the transformation property (\ref{t1}) we consider the example of a point charge $-e$ with mass $m$ bound in the potential $V_{\rm ext}$. Choosing the Coulomb gauge $\nabla \cdot {\bf A} =0$ implies ${\bf A}={\bf A}_{\rm T}$. From Gauss' law $\nabla \cdot {\bf E}=\mu(t)\rho$ we then obtain $A_0= \mu(t)\phi_{\rm Coul}$. Instead of the Coulomb-gauge, we could choose the Poincar\'e-gauge defined by ${\bf x}\cdot {\bf A}({\bf x})=0$. We then obtain ${\bf A}({\bf x})={\bf A}_{\rm T}({\bf x})+\nabla \chi^1$ and $A_0 = \mu(t)\phi_{\rm Coul} -\partial_t\chi^1$ where
\begin{align}\label{chi}
\chi^1({\bf x}) = -\int_0^1 d\lambda\, {\bf x} \cdot {\bf A}_{\rm T}(\lambda {\bf x}).
\end{align}
More generally, we can straightforwardly encode the choice of gauge in a real parameter $\alpha$ such that
\begin{align}\label{vec1}
{\bf A}^\alpha= {\bf A}_{\rm T}+\nabla\chi^\alpha,~~~~A_0^\alpha=\mu(t)\phi_{\rm Coul}-\partial_t \chi^\alpha
\end{align}
where $\chi^\alpha=\alpha \chi^1$ with $\chi^1$ given in Eq.~(\ref{chi}). Note that ${\bf A}_{\rm T}$ is gauge-invariant \cite{cohen-tannoudji_photons_1997}, that is, ${\bf A}_{\rm T}^\alpha={\bf A}_{\rm T}^{\alpha'}$ for all $\alpha$ and $\alpha'$, because $\nabla\chi^\alpha$ is necessarily longitudinal; $\nabla\times \nabla\chi^\alpha \equiv {\bf 0}$.

If we now apply the canonical procedure to derive the Hamiltonian from the Lagrangian in Eq.~(\ref{lag2}) we obtain  
\begin{align}\label{hal}
H^\alpha(t) =& {1\over 2m}\left[{\bf p}+e\mu(t){\bf A}^\alpha({\bf r})\right]^2 + V_{\rm ext}+V_{\rm self}(t) \nonumber \\ &+{1\over 2}\int d^3x \left[\left({\bf \Pi}+\mu(t){\bf P}^\alpha_{\rm T}\right)^2+{\bf B}^2\right]
\end{align}
where
\begin{align}
V_{\rm self}(t)= {\mu(t)^2\over 2 }\int d^3 x \rho\phi_{\rm Coul}
\end{align}
is the infinite Coulomb self-energy of the charge, which is usually taken as renormalising the bare mass and is then ignored, and where
\begin{align}\label{pol}
P_{{\rm T},i}^\alpha = -e\alpha\int_0^1 d\lambda \, r_j\delta_{ij}^{\rm T}({\bf x}-\lambda {\bf r})
\end{align}
is the $\alpha$-gauge transverse polarisation. 
Interpreted as Schr\"odinger picture quantum operators the Hamiltonians of different gauges are non-equivalent being unitarily related by a generalised time-dependent Power-Zienau-Woolley transformation as
\begin{align}\label{hrel}
H^{\alpha'}(t) = R_{\alpha\alpha'}(t)H^\alpha(t)R_{\alpha' \alpha}(t),
\end{align}
where
\begin{align}\label{Ralph}
R_{\alpha\alpha'}(t) = \exp \left[i(\alpha-\alpha')\mu(t)\int d^3 x\, {\bf P}_{\rm T}^1\cdot {\bf A}_{\rm T}\right].
\end{align}
The non-equivalence of the Hamiltonians for distinct values of $\alpha$ follows from Eq.~(\ref{hrel}), which shows that $H^{\alpha'}(t) \neq R_{\alpha\alpha'}(t)H^\alpha(t)R_{\alpha' \alpha}(t) + i{\dot R}_{\alpha\alpha'}(t)R_{\alpha'\alpha}(t)$, where the right-hand-side is equivalent to $H^\alpha(t)$.

Equation (\ref{hal}) gives the $\alpha$-gauge Hamiltonian with time-dependent coupling and no approximations have been made in its derivation, except the use of $\mu(t)$ as a model for the control subsystem. Note that the canonical coordinate of the electromagnetic subsystem is the transverse vector potential ${\bf A}_{\rm T}$, which is manifestly gauge-invariant. The $\alpha$-gauge vector potential ${\bf A}^\alpha$ appearing in Eq.~(\ref{hal}) is specified as a function of ${\bf A}_{\rm T}$ given uniquely by Eqs. (\ref{chi}) and (\ref{vec1}). In particular, these equations together with Eq.~(\ref{pol}) imply that $e{\bf A}^\alpha({\bf r})$ can be written
\begin{align}
e{\bf A}^\alpha({\bf r}) = e{\bf A}_{\rm T}({\bf r}) +\alpha\nabla_{\bf r} \int d^3x \, {\bf P}_{\rm T}^1\cdot {\bf A}_{\rm T}.
\end{align}
It is common to perform the electric dipole approximation (EDA) ${\bf A}_{\rm T}({\bf r})\approx {\bf A}_{\rm T}({\bf 0})$ and $P_{{\rm T},i}^1({\bf x})\approx  -er_j\delta_{ij}^{\rm T}({\bf x})$, which requires the resonant wavelengths to be long compared with the spatial extent of the material system set by $V_{\rm ext}$. It is also common to neglect the infinite self-energy of the charge. One then obtains the Hamiltonian
\begin{align}\label{hamalph}
H^\alpha(t) =& {1\over 2m}\left[{\bf p}+e\mu(t)(1-\alpha){\bf A}_{\rm T}({\bf 0})\right]^2+ V_{\rm ext} \nonumber \\ &+{1\over 2}\int d^3x \left[\left({\bf \Pi}({\bf x})+\alpha\mu(t){\bf P}_{\rm T}^1({\bf x})\right)^2+{\bf B}({\bf x})^2\right].
\end{align}
The choice $\alpha=0$ provides the time-dependent version of the well-known ``${\bf p}\cdot {\bf A}$"-interaction of the Coulomb-gauge, while the choice $\alpha=1$ likewise provides the time-dependent version of the well-known ``$-e{\bf r}\cdot {\bf \Pi}$"-interaction of the Poincar\'e gauge. Both of these interaction forms are commonly found within the literature. The Hamiltonians of different gauges continue to be non-equivalent and unitarily related as in Eq.~(\ref{hrel}) where now
\begin{align}
R_{\alpha\alpha'}(t)=\exp\left[-i(\alpha-\alpha')e\mu(t){\bf r}\cdot {\bf A}_{\rm T}({\bf 0})\right],
\end{align}
which is simply the dipole approximation of Eq. (\ref{Ralph}).

The canonical formalism explains why the non-equivalence of the $H^\alpha(t)$ occurs; in different gauges the theoretical quantum subsystems are defined in terms of {\em different} gauge-invariant observables. In the $\alpha$-gauge the field canonical momentum is ${\bf \Pi}=-{\bf E}_{\rm T}-\alpha{\bf P}_{\rm T}^1$. The Coulomb and multipolar gauges are special cases with ${\bf \Pi}=-{\bf E}_{\rm T}$ and ${\bf \Pi}=-{\bf D}_{\rm T}$ respectively, both of which are gauge-invariant observables. Since the ``light" and ``matter" subsystems are defined using the canonical operator sets $\{{\bf A}_{\rm T},{\bf \Pi}\}$ and $\{{\bf r},{\bf p}\}$, they can also only be specified {\em relative} to a choice of gauge. The interaction being externally controllable constitutes a different physical assumption when imposed on different physical subsystems that are defined relative to different gauges \cite{stokes_implications_2020}. Thus, each $H^\alpha(t)$ describes a different experimental protocol, in which a different interaction is being controlled. This will be demonstrated directly by way of example in Sec.~\ref{gih}.

Presented with an experiment that we are asked to model using a time-dependent coupling, we possess an infinity of non-equivalent models $H^\alpha(t)$ which for each different value of $\alpha$, we know to model a different experimental protocol. Without an argument to choose between the available models an ambiguity is encountered. Determining the correct model may be difficult, because as we shall see, the theoretical subsystems differ between gauges only in their description of virtual processes. In weak-coupling regimes involving sufficiently adiabatic interactions the ``ambiguity" described above is unproblematic, because its consequences are usually negligible in practical calculations. This is no longer the case in sufficiently strong-coupling non-adiabatic regimes where, as is apparent in Eq.~(\ref{hamtim}) below, $\alpha$-dependent components of the interaction $V^\alpha(t)$ are not negligible.

\section{Toy model}\label{hd}

Time-dependent interactions between subsystems arise in many and diverse areas of physics. Here we consider a very simple light-matter model. This serves to clearly determine the situations within which we can expect the gauge non-equivalence of models that result from assuming a time-dependent coupling to become significant. We will see that it becomes significant in the description of so-called virtual processes, which typically become increasingly important with increasing coupling-strength.

\subsection{Time-dependent Hamiltonian}

We suppose that a charge $-e$ is confined in all spatial dimensions except the direction ${\bm \varepsilon}$ of the polarisation of a single cavity-mode, in which it is bound harmonically. The position operator is ${\bf r}=r{\bm \varepsilon}$ and the conjugate momentum is ${\bf p}=p{\bm \varepsilon}$.~The material canonical commutation relation is $[r,p]=i$. The field canonical commutation relation is, in the general case, given by
\begin{align}\label{comtot}
[A_i({\bf x}),\Pi_j({\bf x}')]=i\delta_{ij}^{\rm T}({\bf x}-{\bf x}').
\end{align}
Discretising the modes within a cavity volume $v$ the fields can be expanded in terms of photon creation and annihilation operators. Restricting the fields to a single mode ${\bf k}\lambda$ then gives
\begin{align}
&{\bf A}_{\rm T}({\bf x}) =  g {\bm \varepsilon} \left(a^\dagger {\rm e}^{-{\rm i}{\bf k}\cdot {\bf x}}+a {\rm e}^{{\rm i}{\bf k}\cdot {\bf x}}\right),\label{consm}\\ &{\bf \Pi}({\bf x}) ={\rm i} \omega g {\bm \varepsilon} \left(a^\dagger {\rm e}^{-{\rm i}{\bf k}\cdot {\bf x}}-a {\rm e}^{{\rm i}{\bf k}\cdot {\bf x}}\right),\label{consm2}
\end{align}
where $g=1/\sqrt{2\omega v}$, $\omega=|{\bf k}|$, ${\bm \varepsilon} \equiv {\bm \varepsilon}_{{\bf k}\lambda}$ is orthogonal to ${\bf k}$, and $a \equiv a_{{\bf k}\lambda}$ with $[a,a^\dagger]=1$. Equations (\ref{consm}) and (\ref{consm2}) imply that the cavity canonical operators now satisfy the commutation relation 
\begin{align}\label{tcom2}
[A_{{\rm T},i}({\bf x}),\Pi_j({\bf x}')]={{\rm i}\varepsilon_i \varepsilon_j \over v}\cos \left[{\bf k}\cdot ({\bf x}-{\bf x}')\right].
\end{align}

In the dipole approximated Hamiltonian of Eq.~(\ref{hamalph}) the fields are evaluated at the dipolar position ${\bf 0}$. The Hamiltonian can therefore be expressed entirely in terms of the cavity variables $A={\bm \varepsilon}\cdot {\bf A}_{\rm T}({\bf 0})$ and $\Pi={\bm \varepsilon}\cdot {\bf \Pi}({\bf 0})$. According to Eqs.~(\ref{consm}) and (\ref{consm2}), the commutator of these variables is $[A, \Pi]=i/ v$. Comparing this commutator, or the commutator in Eq.~(\ref{tcom2}), with Eq.~(\ref{comtot}), reveals that the transverse delta-function is given within the single-mode approximation by 
\begin{align}
\delta_{ij}^{\rm T}({\bf 0}) = \int {d^3k \over (2\pi)^3} \sum_\lambda \varepsilon_{\lambda,i} \varepsilon_{\lambda, j}  \longrightarrow {1\over v} \varepsilon_i \varepsilon_j.
\end{align}
It follows that the polarisation self-energy term in the Hamiltonian becomes within the EDA and single-mode approximations
\begin{align}
{1\over 2}\int d^3 x {\bf P}_{\rm T}({\bf x})^2 = {e^2\over 2}r_ir_j\delta_{ij}({\bf 0}) \longrightarrow {e^2\over 2v}r^2.
\end{align}
The dipole and single-mode approximations have no bearing on gauge invariance or non-invariance, because whether they are performed or not, the Hamiltonians $H^\alpha(t)$ are equivalent if and only if ${\dot \mu}=0$. The approximations are used here to enable a simple and transparent example.

Material bosonic ladder operators can be defined as $b =\sqrt{1/2m\omega_m} (m \omega_m r +ip)$ with $[b,b^\dagger]=1$. The Hamiltonian in Eq.~(\ref{hamalph}) can now be written
\begin{align}\label{full}
H^\alpha = H_0+V^\alpha
\end{align}
where $H_0=\omega(a^\dagger a+1/2)+\omega_m(b^\dagger b +1/2)$ and
\begin{align}\label{hamtim}
V^\alpha=\,&{\eta(t)^2\omega \over 4}\left[(1-\alpha)^2(a^\dagger+a)^2 + \delta \alpha^2(b^\dagger+b)^2\right] \nonumber \\ &+ iu_\alpha^-(t)(ab^\dagger - a^\dagger b) + iu_\alpha^+(t)(a^\dagger b^\dagger - ab)
\end{align}
with $\delta =\omega/\omega_m$ and
\begin{align}
\eta(t) &=\eta\mu(t) = {e\mu(t)\over\omega\sqrt{mv}},\\
u_\alpha^\pm(t)&={1\over 2}\eta(t)\omega_m\sqrt{\delta}[(1-\alpha)\mp \delta\alpha].
\end{align}

\subsection{Bare-energy conservation, and $\alpha$-independent predictions}\label{virt}


In QED material systems are often interpreted as surrounded by a cloud of virtual photons \cite{cohen-tannoudji_photons_1997,cohen-tannoudji_atom-photon_2010,compagno_atom-field_1995,passante_cloud_1985,compagno_interference_1985,persico_time_1987}. Two examples of virtual processes are those described by the terms $ab$ and $a^\dagger b^\dagger$ having a coupling strength $u_\alpha^+(t)$ in Eq.~(\ref{hamtim}), which are number non-conserving and so do not commute with $H_0$. 
Inspection of Eq.~(\ref{hamtim}) reveals directly that different subsystem divisions only differ in their description of virtual processes. All number non-conserving terms in Eq. (\ref{hamtim}) are $\alpha$-dependent, whereas the remaining number-conserving part is $\alpha$-independent at resonance, which is precisely when this term conserves the bare energy, because
\begin{align}\label{cons}
{i\over 2}\omega_m\eta(t)[H_0,(ab^\dagger-a^\dagger b)]=0.
\end{align}

Thus, despite the $\alpha$-dependence of the subsystems themselves, within the approximation of retaining only the interaction terms that conserve $H_0$, all $\alpha$-dependence drops out of the theory. In this case the bare vacuum coincides with the Hamiltonian ground state. This approximation is valid in the traditional regime of weakly-coupled nearly-resonant systems, a regime that can be understood as gauge-nonrelativistic \cite{stokes_implications_2020}. Therein, one can pretend that the quantum subsystems,``light" and ``matter", are ostensibly unique, i.e., not gauge-relative. In truth, this is not the case, and the pretence cannot be sustained if the required approximation is not valid. Therefore, for ultrastrong and fast interactions, when assuming a time-dependent coupling it must be determined which gauges describe which experimental protocols and arrangements. In what follows we verify that the correct gauge to use will generally depend on the microscopic arrangement being considered. No one gauge is universally correct.

\begin{figure}[t]
\includegraphics[width=0.9\linewidth, height=7cm, keepaspectratio]{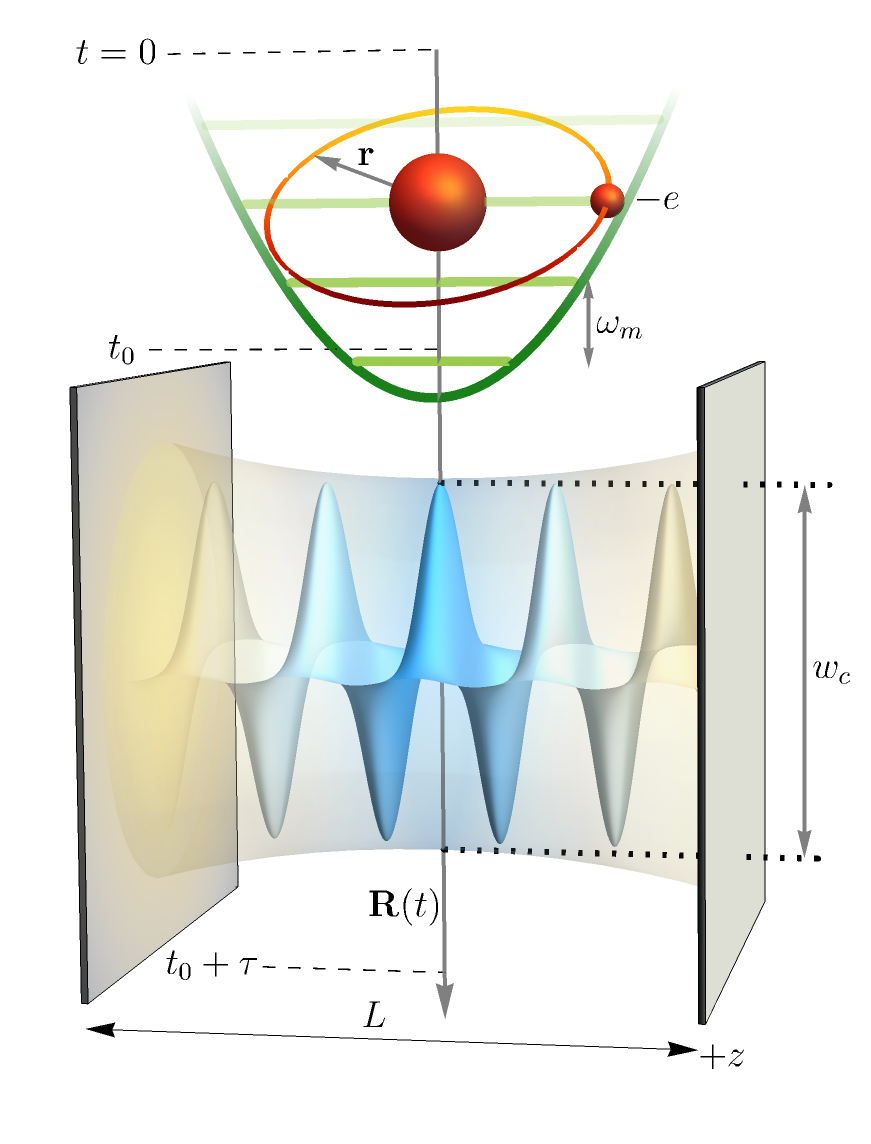}\vspace*{-4mm}
\caption{A cavity of length $L$ supporting standing waves in the $z$-direction and a Gaussian perpendicular mode profile with waist $w_c$ is depicted, along with a dipole $-er$ oscillating with frequency $\omega_m$. At $t=0$ the cavity and dipole are non-interacting. The dipole follows a classical trajectory ${\bf R}(t)$ through the cavity, entering the cavity at $t_0$ and exiting at $t_0+\tau$. The Hamiltonian for this system is derived in Appendix C and can be realised using a time-dependent coupling $\phi({\bf R}(t))=:\mu(t)$ as in Eq.~(\ref{h4}).}\label{pic}
\end{figure}
Ultrastrong and fast interactions are now of major importance \cite{kockum_ultrastrong_2019,forn-diaz_ultrastrong_2019}. For such interactions a model corresponding to $\alpha=0$ or $\alpha=1$, which are both commonly chosen gauges in light-matter theory, will not generally produce even qualitatively accurate predictions if the underlying physics of the system is more correctly described by an interaction corresponding, for example, to $\alpha \sim \alpha_g= 1/(1+\delta)$, for which $u^+_\alpha(t)$ vanishes identically. In fact, as we shall show, even conventional gauges $\alpha=0$ and $\alpha=1$ generally give significantly different physical predictions when the coupling $\eta(t)$ is ultrastrong and ultra-fast, because the two models possess different dependencies on the underlying model parameters.

We remark that anharmonic material systems may also be considered. The $\alpha$-independence of predictions for processes conserving the non-interacting part of a Hamiltonian is a completely general result within scattering theory \cite{cohen-tannoudji_photons_1997,craig_molecular_1998,woolley_gauge_2000,stokes_implications_2020}. However, anharmonic matter does not generally admit a simple analytic treatment at the level of the model Hamiltonian. An exception is when the material system is sufficiently anharmonic that a two-level truncation can be made. In general this will break the gauge-invariance of the theory and so must be performed within a gauge in which the truncation is found to be accurate for the properties of interest \cite{de_bernardis_breakdown_2018,stokes_gauge_2019,di_stefano_resolution_2019,roth_optimal_2019}.  

\section{Example: Uniform motion through a cavity}\label{cavity}

\subsection{Time-dependent Hamiltonian}\label{tdh}

The variation of $\mu(t)$ could be interpreted as a model for the motion of an external potential $V_{\rm ext}$, which moves in and out of contact with the electromagnetic fields. As we noted in Sec.~\ref{noneqlag}, the system responsible for the potential can be called the {\em control system}, which in a more complete description would be included explicitly via additional dynamical position and momentum variables ${\bf R}$ and ${\bf K}$. 
To show how the non-equivalence of the Hamiltonians $H^\alpha(t)$ results from an approximation, we consider a hydrogen atom consisting of a charge $+e$ with mass $m_2$ at ${\bf r}_2$ and a charge $-e$ with mass $m_1$ at ${\bf r}_1$. The charge and current densities are $\rho({\bf x}) = e[\delta({\bf x}-{\bf r}_1) - \delta({\bf x}-{\bf r}_2)]$ and ${\bf J}({\bf x})=e[{\dot {\bf r}}_1\delta({\bf x}-{\bf r}_1)-{\dot {\bf r}}_2\delta({\bf x}-{\bf r}_2)]$. Within the EDA
\begin{align}
\rho({\bf x})&=e{\bf r}\cdot \nabla\delta({\bf x}-{\bf R}),\label{dens}\\
{\bf J}({\bf x}) &=-e{\dot {\bf r}}\delta({\bf x}-{\bf R}) +e{\dot {\bf R}}({\bf r}\cdot \nabla)\delta({\bf x}-{\bf R}),\label{curr}
\end{align}
where ${\bf r}={\bf r}_1-{\bf r}_2$ is the relative position between the charges and ${\bf R}=(m_1{\bf r}_1+m_2{\bf r}_2)/(m_1+m_2)$ is the position of the centre-of-mass. The second term on the right-hand-side of Eq.~(\ref{curr}) ensures the conservation of charge $\partial_\mu j^\mu=0$, and also ensures that the dipole's interaction with the electric field induced by the atomic motion in the lab frame is properly included. In particular, in the multipolar-gauge this term correctly ensures the presence of the R\"ontgen interaction. The current can be obtained from the non-relativistic transformation $\rho=\rho'$, ${\bf J}={\bf J}'+{\dot {\bf R}}\rho'$, which relates the (primed) atomic rest frame to the (unprimed) lab frame in which ${\bf R}={\dot {\bf R}}t$ (up to a constant initial position) with ${\dot {\bf R}}\neq {\bf 0}$. The $\alpha$-gauge polarisation field is
\begin{align}\label{pola}
P_{{\rm T},i}^\alpha({\bf x}) = -e\alpha r_j\delta_{ij}^{\rm T}({\bf x}-{\bf R})
\end{align}
and the associated magnetisation field ${\bf M}^\alpha$ is such that $\nabla \times {\bf M}^\alpha = {\bf J}_{\rm T}-{\dot {\bf P}}^\alpha_{\rm T}$. In the multipolar gauge $\alpha=1$ we obtain the expected multipolar expressions. In particular, $\nabla \times {\bf M}^1({\bf x})  = -e\nabla \times [{\bf r}\times {\dot {\bf R}}\delta({\bf x}-{\bf R})]$.

Eqs.~(\ref{dens}), (\ref{curr}) and (\ref{pola}) can be used within the Lagrangian of Eq. (\ref{Lag}) and the $\alpha$-gauge Hamiltonian can be derived with ${\bf r}$, ${\bf R}$, and ${\bf A}_{\rm T}$ as canonical coordinates. Details are given in Appendix~B. If we approximate the centre-of-mass position ${\bf R}$ as externally prescribed within the Hamiltonian then we obtain a bipartite quantum system and the Hamiltonians of different gauges become non-equivalent, being given by
\begin{align}\label{hamalph2}
H^\alpha(t) =&{1\over 2m}[{\bf p}+e(1-\alpha){\bf A}_{\rm T}({\bf R(t)})]^2 + V({\bf r})  \nonumber \\ &+ {1\over 2}\int d^3x \left[({\bf \Pi}+{\bf P}^\alpha_{\rm T}(t))^2+{\bf B}^2\right],
\end{align}
where ${\bf P}_{\rm T}^\alpha(t)$ is explicitly time-dependent due to its dependence on ${\bf R}$. This expression clearly has the same structure as Eq.~(\ref{hal}) and its dipole approximation, Eq.~(\ref{hamalph}), which were obtained by assuming a time-dependent coupling.

To progress further, in Appendix~C we specialise the above Hamiltonian to the case of a Fabry-Perot Gaussian cavity mode with mirrors orthogonal to the $z$-direction, as depicted in Fig. \ref{pic}. Assuming as before that ${\bf r}=r{\bm \varepsilon}$ and ${\bf p}=p{\bm \varepsilon}$, and that the atomic potential energy $V(r)$ is harmonic, the Hamiltonian reads as
\begin{align}\label{h4}
H^\alpha(t)=& {1\over 2m}[p + e(1-\alpha) A_{\rm T}({\bf R}(t))]^2 + {m\omega_m^2 \over 2}r^2 \nonumber \\ & +e^2{\alpha^2 r^2 \over 2v}|\phi({\bf R}(t))|^2 - e\alpha r \Pi({\bf R}(t)) \nonumber \\ &+ \omega\left(a^\dagger a +{1\over 2}\right),
\end{align}
where $\phi({\bf x})=e^{ikz}e^{-(x^2+y^2)/w_c^2}$ is a Gaussian mode envelope, with $w_c$ the Gaussian beam waist. We assume the path of the atom to be in the $xy$-plane such that ${\hat {\bf z}}\cdot {\bf R}(t)=0$. The cavity canonical operators are given by
\begin{align}
A_{\rm T}(t,{\bf x}) &=  {1\over \sqrt{2\omega v}} [\phi^*({\bf x}) a^\dagger (t) +\phi({\bf x}) a(t)], \label{mod1a} \\
\Pi(t,{\bf x}) &= i \sqrt{\omega \over 2v} [\phi^*({\bf x}) a^\dagger (t) -\phi({\bf x}) a(t)].\label{mod2a}
\end{align}
Letting $\phi({\bf R}(t))=\mu(t)$, we see that the Hamiltonian in Eq.~(\ref{h4}) is identical to that defined in Eq.~(\ref{full}), which was in turn obtained from Eq.~(\ref{hal}). 
This shows that the non-equivalence of the models $H_\alpha(t)$ for a single atom obtained via the time-dependent coupling method can indeed be understood as a consequence of approximating the centre of mass motion (the control subsystem) as an external subsystem. 
\begin{figure}[t]
\begin{minipage}{\columnwidth}
\begin{center}
\vspace*{-3mm}
\hspace*{-0.2cm}\includegraphics[scale=0.44]{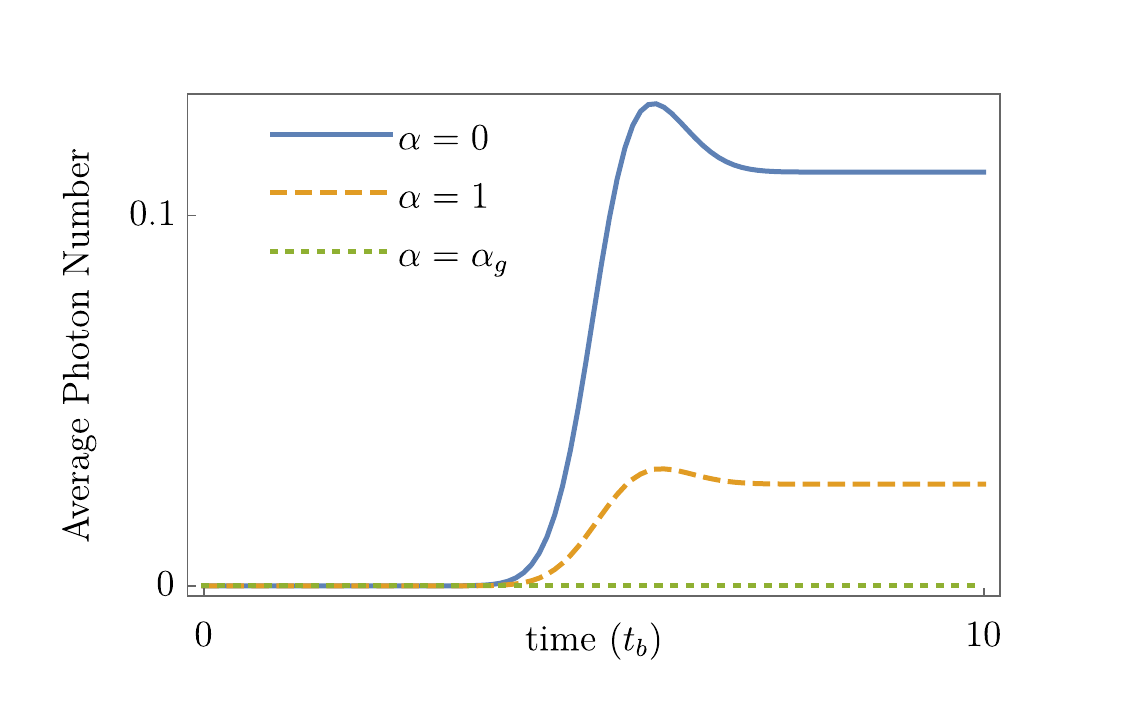}
\vspace*{-4mm}
\caption{$\eta=1$ and $\delta=1/2$. The average number of photons found using $H^\alpha(t)$ is plotted with time in units of the beam transit time $t_b=w_c/\nu$ assuming an initial state $\ket{0,0}$. The beam waist is $w_c=20\mu$m, $\omega_m$ is chosen in the microwave regime (energy $\sim10\mu$eV) and $\nu = 10^{-3}c$ where $c$ is the speed of light. The final values are given where curves become straight, and are clearly different for different $\alpha$.}\label{ph_g}
\end{center}
\end{minipage}
\end{figure}

Uniform motion of the dipole in and out of the cavity is described by a Gaussian function $\mu(t)$. Significant $\alpha$-dependence of final predictions occurs when the interaction time $\tau\sim w_c/\nu$ ($\nu={\dot R}$) is comparable to the cycle time $1/\omega_m$. In the case of a micro-cavity with $w_c=20~\mu$m and $\omega_m$ in the microwave regime, this requires $\nu\sim 10^{-3}c$, which is non-relativistic. We assume that the system is initially non-interacting ($\mu(0)=0$) and starts in the ground state $\ket{0,0}$. The interaction is switched on at time $t_0>0$, and switched-off at $t_0+\tau>t_0$. Thus, at the preparation and measurement stages the definitions of the quantum subsystems are unique. In Fig.~\ref{ph_g} the number of cavity photons is plotted as a function of time with $w_c\omega_m/\nu\sim 1$, $\eta= 1$, and $\alpha=0,\,1,\,\alpha_g$. The three gauges give different residual photon populations within the cavity after the interaction has ceased, consistent with the suggestion of the energy-time uncertainty relation. For longer and slower interaction switching and weaker couplings than we have shown all photon populations return to zero independent of $\alpha$. In contrast, when the interaction switching is on the order of a bare cycle and the coupling is sufficiently strong, there is a significant probability that virtual photons created near the beginning of the switch-off are not reabsorbed before the atom has exited the cavity. They therefore detach and are left behind, remaining inside the cavity. This cannot occur however, for values of $\alpha\sim \alpha_g$ for which ground state photons are not explicit.

\subsection{Gauge-invariant class of approximate Hamiltonians}\label{gih}

In the case of a sufficiently simple moving bound-charge system, as considered above, a gauge-invariant set of Hamiltonians can be derived. This can only be achieved by starting with an explicit model for the control system and requires making the approximation that this control is external at the Lagrangian level, rather than at the Hamiltonian level. The procedure is {\em not} generally equivalent to assuming a time-dependent coupling.

In the case of the hydrogen atom within the EDA, 
we have three subsystems with position coordinates ${\bf r}$, ${\bf R}$, and ${\bf A}_{\rm T}$. The control subsystem with coordinate ${\bf R}$ can be treated as external at the Lagrangian level, such that it becomes a pre-prescribed dynamical vector, ${\bf R}(t)$, {\em prior} to the transition to the canonical formalism. Now only ${\bf r}$ and ${\bf A}_{\rm T}$ remain as dynamical variables. The resulting $\alpha$-gauge Hamiltonian, denoted ${\tilde H}^\alpha(t)$, is given by
\begin{align}\label{h45}
{\tilde H}^\alpha(t) = &H^\alpha(t)+e{\dot {\bf R}}\cdot\left[ ({\bf r}\cdot \nabla){\bf A}_{\rm T}({\bf R})\right]\nonumber \\ &-e\alpha {\dot {\bf R}}\cdot \nabla[{\bf r}\cdot {\bf A}_{\rm T}({\bf R})]
\end{align}
where $H^\alpha(t)$ is given in Eq.~(\ref{hamalph2}). Full details of the derivation are given in Appendix~B. Unlike the Hamiltonians $H^\alpha(t)$, as Schr\"odinger picture operators the Hamiltonians ${\tilde H}^\alpha(t)$ satisfy 
\begin{align}
{\tilde H}^\alpha(t) = R_{0\alpha}{\tilde H}^0(t) R_{0\alpha}^\dagger +i{\dot R}_{0\alpha}R_{0\alpha}^\dagger
\end{align}
such that Hamiltonians belonging to distinct gauges are equivalent. It is instructive to consider the multipolar-gauge example $(\alpha=1)$;
\begin{align}\label{htil1}
{\tilde H}^1(t) =& {{\bf p}^2\over 2m}+V({\bf r}) + {1\over 2}\int d^3 x \left[({\bf \Pi}+{\bf P}^1_{\rm T})^2+{\bf B}^2\right]\nonumber \\ &+e{\bf r}\cdot[{\dot {\bf R}}\times {\bf B}({\bf R})].
\end{align}
The final term in this expression describes the R\"ontgen interaction in which the dipole experiences an effective electric field ${\dot {\bf R}}\times {\bf B}({\bf R})$ due to the gross motion in the lab frame \cite{craig_molecular_1998,lembessis_theory_1993,boussiakou_quantum_2002}. Such an interaction also appears in the complete Hamiltonian derived by keeping ${\bf R}$ as a dynamical variable where it manifests via a non-mechanical canonical momentum ${\bf K} = M{\dot {\bf R}} +e{\bf r} \times {\bf B}({\bf R})$. The R\"ontgen interaction term is lost when ${\bf R}={\dot {\bf R}}t$ is prescribed as external within the complete Hamiltonian, which results in Eq.~(\ref{hamalph2}).


When using ${\tilde H}^\alpha(t)$ any value of $\alpha$ can be chosen and the final predictions will necessarily be $\alpha$-independent (gauge-invariant). Let us therefore suppose that these predictions are ``correct" (albeit approximate). It follows that the fixed values of $\alpha$ for which $H^\alpha(t)={\tilde H}^\alpha(t)$ are those allowed in order to obtain ``correct" results using $H^\alpha(t)$.
\begin{figure}[t]
\begin{minipage}{\columnwidth}
\begin{center}
\vspace*{-3mm}
\hspace*{-0.2cm}\includegraphics[scale=0.44]{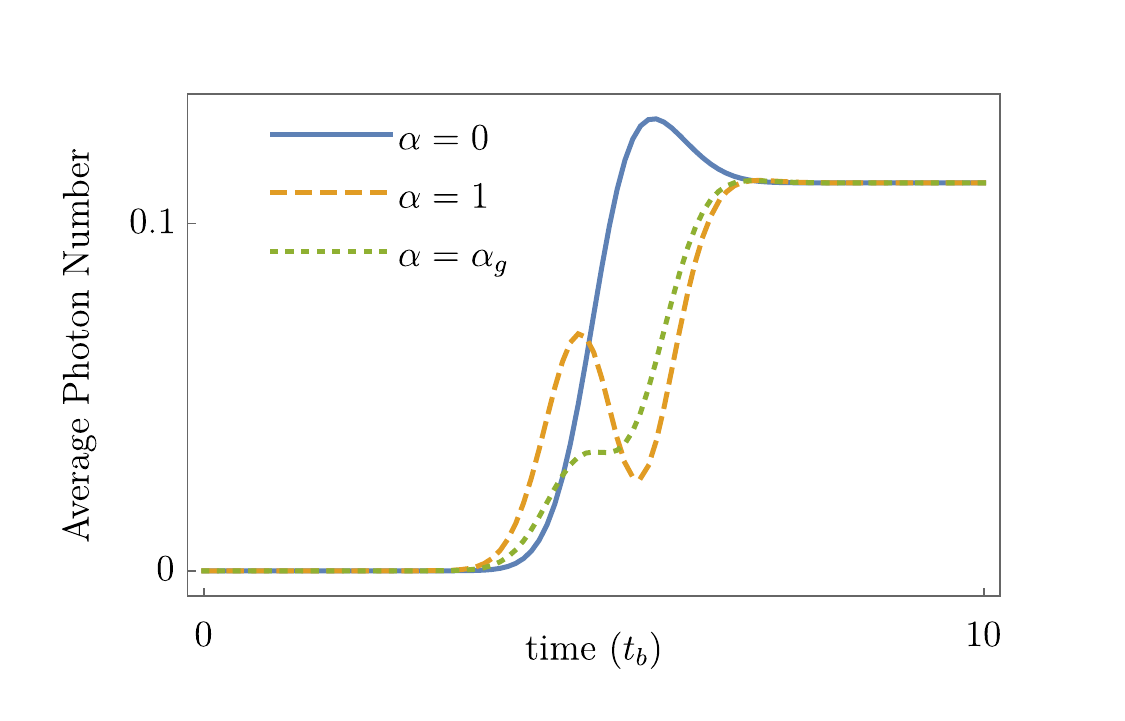}
\vspace*{-4mm}
\caption{All parameters are as in Fig. \ref{ph_g}. The average number of photons is plotted with time in units of $t_b=w_c/\nu$ assuming an initial state $\ket{0,0}$. The dynamics are generated by ${\tilde H}^\alpha(t)$ and we have assumed $\cos\theta = 0$. As expected, during the interaction window the average photon number differs between gauges. However, in contrast to Fig. \ref{ph_g} all gauges predict the same final value. Because we have chosen $\cos\theta=0$, the final value coincides with that predicted by $H^0(t)$. It is therefore identical to the final value of the $\alpha=0$ curve in Fig. \ref{ph_g}}\label{ph_5}
\end{center}
\end{minipage}
\end{figure}
Taking the fields given by Eqs.~(\ref{mod1a}) and (\ref{mod2a}), we assume that ${\bf R}(t)=(h-\nu t){\hat {\bf x}}$, implying uniform motion ${\dot {\bf R}}=-\nu{\hat {\bf x}}$ from an initial position $h{\hat {\bf x}}$ outside of the cavity. Under these conditions the difference ${\tilde H}^\alpha(t)-H^\alpha(t)$ is given by
\begin{align}\label{exterm}
&e{\dot {\bf R}}\cdot\left[ ({\bf r}\cdot \nabla){\bf A}_{\rm T}({\bf R})\right]-e\alpha {\dot {\bf R}}\cdot \nabla[{\bf r}\cdot {\bf A}_{\rm T}({\bf R})]\nonumber \\
=&{e\nu\over \sqrt{2\omega v}}\phi({\bf R})\varepsilon_i r_j \nonumber \\ & \times \bigg[ i\omega {\hat x}_i {\hat z}_j(a^\dagger -a) -{2R\over w_c^2}\left(\alpha \delta_{ij}-{\hat x}_i{\hat x}_j\right)\left(a^\dagger +a\right) \bigg].
\end{align}
It is straightforward to verify that for $\alpha=1$ the right-hand-side of this expression coincides with $e{\bf r}\cdot[{\dot {\bf R}}\times (\nabla \times {\bf A}_{\rm T}({\bf R}))]$ as required [cf. Eq~(\ref{htil1})]. Since the coefficient of $a^\dagger -a$ on the right-hand-side is $\alpha$-independent, there is no choice of $\alpha$ for which $H^\alpha(t)={\tilde H}^\alpha(t)$ in general. However, if we make the simplifying assumption that ${\bf r}=r{\bm \varepsilon}$ [giving $H^\alpha(t)$ as in Eq.~(\ref{h4})], then we obtain
\begin{align}\label{diff}
{\tilde H}^\alpha(t)-H^\alpha(t)=-e{\dot \mu}(t)rA_{\rm T}({\bf 0})\left[\alpha-\cos^2\theta\right]
\end{align}
where $\cos\theta={\bm \varepsilon}\cdot {\hat {\bf x}}$. Notice that if the switching ${\dot \mu}$ is sufficiently slow then ${\tilde H}^\alpha(t)=H^\alpha(t)$ independent of $\alpha$, whereas if ${\dot \mu}$ is sufficiently fast then the predictions obtained from different $H^\alpha(t)$ may become appreciably different. In contrast, the predictions obtained from ${\tilde H}^\alpha(t)$ are always $\alpha$-independent (gauge-invariant). The correct value of $\alpha$ to choose within $H^\alpha(t)$ is the value solving the equation $H^\alpha(t)={\tilde H}^\alpha(t)$. This value depends on the orientation of the mode polarisation and dipole moment ${\bm \varepsilon}$ relative to the direction of motion ${\hat {\bf x}}$. In other words, the correct value of $\alpha$ to choose when employing the time-dependent coupling method, depends on the microscopic arrangement of the system (the microscopic context).

Consider the arrangements $\theta=\pm\pi/2$ (${\bm \varepsilon}$ and ${\hat {\bf x}}$ orthogonal). From Eq.~(\ref{diff}) we have ${\tilde H}^0(t)|_{\theta=\pm\pi/2} = H^0(t)$ whereas ${\tilde H}^\alpha(t)|_{\theta=\pm\pi/2} \neq H^\alpha(t)$ for $\alpha\neq 0$. Therefore, to model these arrangements the correct value of $\alpha$ to choose when using $H^\alpha(t)$ is $\alpha=0$. Any other value will yield incorrect predictions as determined by comparison with the gauge-invariant predictions provided by ${\tilde H}^\alpha(t)|_{\theta=\pi/2}$. Similarly, for the alternative arrangements $\theta=0,\pi$ (${\bm \varepsilon}$ and ${\hat {\bf x}}$ parallel) we have ${\tilde H}^1(t)|_{\theta=0,\pi} = H^1(t)$ whereas ${\tilde H}^\alpha(t)|_{\theta=0,\pi} \neq H^\alpha(t)$ for $\alpha\neq 1$. Therefore $\alpha=1$ is the correct value to choose when modelling these arrangements using $H^\alpha(t)$. More generally, Eq.~(\ref{diff}) shows that for modelling the arrangement $\theta$ a correct value of $\alpha$ to choose when using $H^\alpha(t)$, is a solution of $\alpha=\cos^2\theta$. 

We have demonstrated that the determination of when the time-dependent coupling method will produce correct results cannot be accomplished without recourse to a more complete description, which yields the constraint $\alpha= \cos^2\theta$. Under this constraint $H^\alpha(t)$ provides a gauge-invariant description, because it coincides with ${\tilde H}^\alpha(t)$ which provides a gauge-invariant description by construction. Under the constraint that $\alpha=\cos^2 \theta$, the parameter $\alpha$ may be thought of as selecting an experimental context rather than a choice of gauge. It follows that it is not the case that the Coulomb-gauge is always correct when using a time-dependent coupling, contradicting Ref.~\cite{di_stefano_resolution_2019}. For example, within the system considered here the Coulomb-gauge will yield the correct description only when $\cos\theta=0$. Subsequent time-dependent gauge-transformation using $R_{0\alpha}(t)$ will of course then yield an equivalent description to $H^0(t)$ as noted in Ref.~\cite{di_stefano_resolution_2019}, but this equivalence class of models is restricted to describing the experimental context $\cos \theta =0$. 


In Fig. \ref{ph_5} the average number of photons is plotted as a function of time found using ${\tilde H}^\alpha(t)$. As expected, the number differs between gauges when $\eta\neq 0$, due to the inherent relativity in the definition of the light quantum subsystem, but in contrast to the predictions obtained using $H^\alpha(t)$ (cf. Fig.~\ref{ph_g}), in Fig.~\ref{ph_5} all gauges predict the same final value.  Since we have chosen $\cos\theta=0$, this final value coincides with that predicted by $H^0(t)$ as given by the curve for $\alpha=0$ in Fig.~\ref{ph_g}. Similarly, if $\cos\theta = 1$ the final value coincides with that predicted by $H^1(t)$, which is given by the $\alpha=1$ curve in Fig.~\ref{ph_g}.

Considering a uniform distribution of random orientations $\theta$ the average of Eq.~(\ref{diff}) is
\begin{align}\label{diff2}
E[{\tilde H}^\alpha(t)-H^\alpha(t)]_\theta=-e{\dot \mu}(t)rA_{\rm T}({\bf 0})\left[\alpha-{1\over 2}\right].
\end{align}
At resonance ($\delta=1$) the Jaynes-Cummings gauge $\alpha_g= 1/(1+\delta)$ now gives the ``correct" value, but the difference $|1/2 - \alpha_g|$ increases as the detuning moves away from resonance.

\section{General time-dependent coupling}\label{general}

A tuneable coupling function could be used to model any time-dependent interaction, such as those realised by addressing specific states of an atomic system \cite{gunter_sub-cycle_2009}, or laser driven systems \cite{cohen-tannoudji_atom-photon_2010}. Switchable interactions are also commonly encountered in superconducting circuits \cite{peropadre_switchable_2010}. It is seldom the case that the subsystem mediating a controllable interaction between two other subsystems will admit a straightforward explicit model of the kind that we have been able to provide for the simple atom-cavity example considered above. It is therefore important to understand more generally the extent to which results obtained from the simple time-dependent coupling method will apply only to a specific experimental context. 
To this end we consider the general coupling function
\begin{align}\label{gen_coup}
\mu(t)=1-{\tanh \left(\frac{s t_0}{2}\right)\sinh ^2\left(\frac{s}{2} \left(t-\frac{\tau }{2}-t_0\right)\right) \over \text{cosh}\left(\frac{s}{2} (t-t_0)\right) \text{cosh}\left(\frac{s}{2} (\tau +t_0-t)\right)}.
\end{align}
This is a smoothed box-function with a maximum of one at $t=t_0+\tau/2$, such that $\mu(t_0)\approx 1/2$, and $\tau$ is roughly the full-width at half maximum. The parameter $s$ controls the smoothness of the switch-on. Through tuning of parameters this general coupling function can take a variety of forms, including close resemblance to a Gaussian, as occurs for uniform atomic motion through a Gaussian cavity. In what follows we determine the dependence on $\alpha$ of the final light and matter properties that result from the dynamics generated by $H^\alpha(t)$ in non-adiabatic strong-coupling regimes.

A naive example of time-dependent coupling comprises instantaneous switching of a constant interaction, but here the free evolution before and after the interaction window does not alter the physical quantities of interest. Predictions for the case of a constant interaction in the ground state $\ket{G}$ of the full Hamiltonian $H^\alpha$ are given in Appendix~D.

\begin{figure}[t]
\begin{minipage}{\columnwidth}
\begin{center}
\hspace*{-4mm}\includegraphics[scale=0.41]{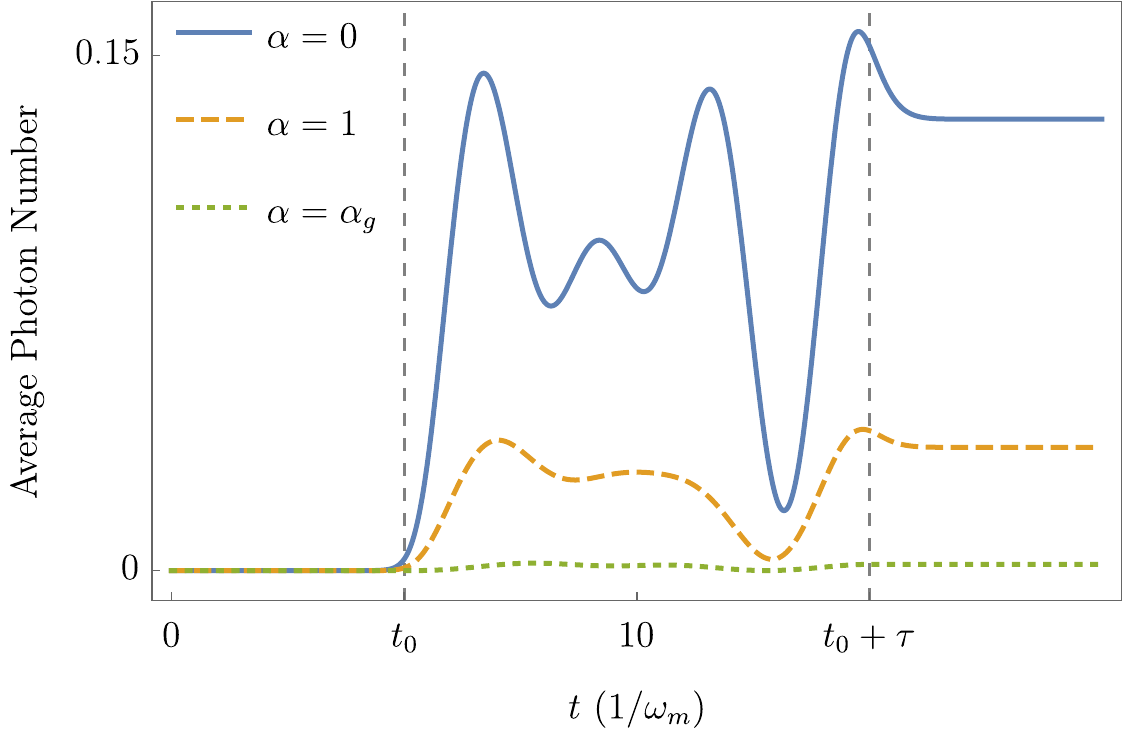}
\vspace*{-3mm}\caption{Starting at $t=0$ in the ground state $\ket{0,0}$ of $H_0=H^\alpha(0)$, the average number of photons $\langle a^\dagger a\rangle_t$ in the cavity is plotted with time. The switch-on function $\mu(t)$ is shown in Fig. \ref{mi2} and is such that the switch-on time is roughly $4/\omega_m$. The remaining parameters are $\eta=1$ and $\delta=1/2$, and $\omega_m$ is in the microwave regime. The final values after the interaction has ceased are given where the curves level-off, and are clearly different for different $\alpha$.}\label{time}
\end{center}
\end{minipage}
\end{figure}

A more realistic interaction switching is smooth, requiring finite time. We therefore use the general coupling function given in Eq.~(\ref{gen_coup}) within the simple Hamiltonian given in Eq.~(\ref{full}). The dynamics of the system are found by numerically solving the closed set of differential equations for correlations of the form $\langle xy \rangle$ where $x,\,y=a,\,a^\dagger,\,b,\,b^\dagger$. For an initial Gaussian state these correlations suffice to completely characterise the final state \cite{olivares_quantum_2012}. We find that significant $\alpha$-dependence of final predictions occurs if the interaction switching time is ultra-fast, i.e., of the order of a bare cycle $\omega^{-1},\,\omega_m^{-1}$, and the coupling is sufficiently strong. For longer switching times predictions from different gauges converge as the interaction is switched-off, such that no differences remain by the end of the protocol. Figure~\ref{time} shows the average number of photons in the cavity as a function of time, when the switching time is roughly $4/\omega_m$ and the system starts in the ground state $\ket{0,0}$ of $H_0=H^\alpha(0)$. Again, both initially and finally there is no ambiguity in the definitions of the light and matter systems, which are uncoupled. Relevant sub-cycle, ultrastrong couplings have already been achieved in cavity QED \cite{gunter_sub-cycle_2009}.

\begin{figure}[t]
\begin{minipage}{\columnwidth}
\begin{center}
\includegraphics[scale=0.43]{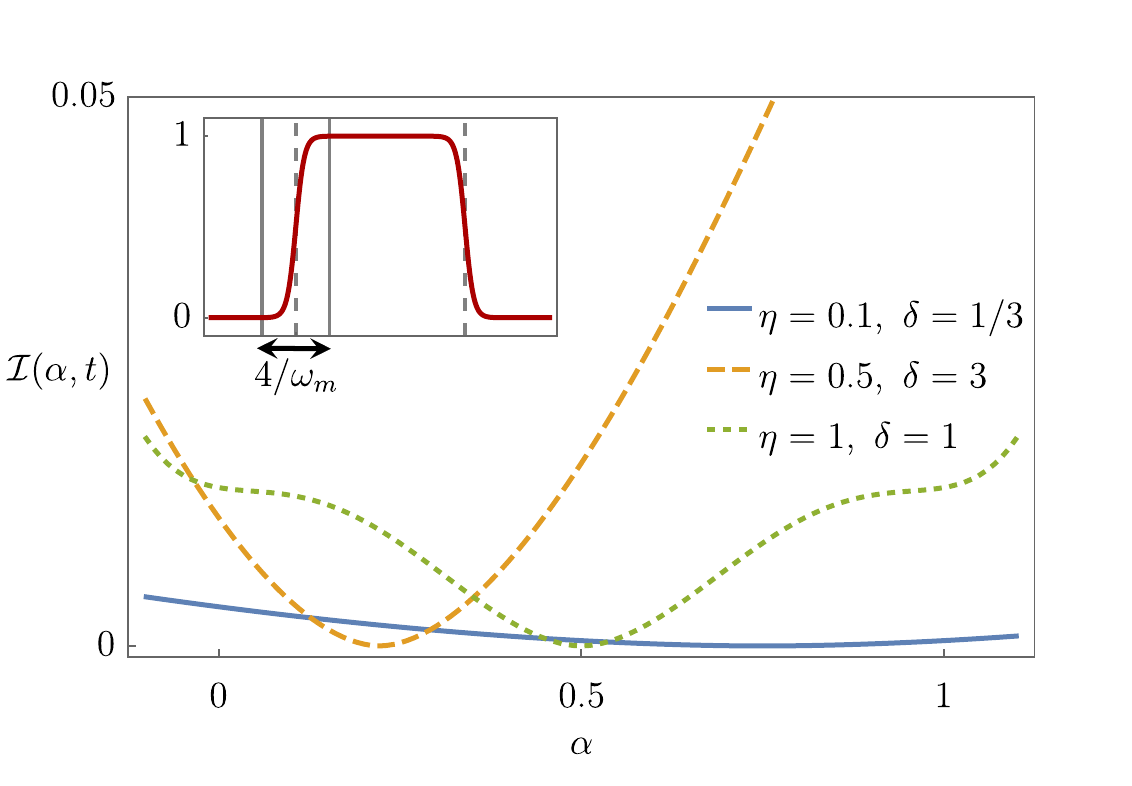}
\caption{Starting at $t=0$ in the ground state $\ket{0,0}$ of $H_0=H^\alpha(0)$, the mutual information ${\cal I}(\alpha)$ at a final time $t$ long after the interaction has been switched-off, is plotted as a function of $\alpha$ for various combinations of $\delta$ and $\eta$. The inset shows the coupling envelope $\mu(t)$ as a function of time. The interaction duration given by the difference in the dashed lines is $\tau=10/\omega_m$ with $\omega_m$ chosen in the optical range. The switch on occurs at roughly $t_0=\tau/2$ and the chosen value of $s$ gives a switching time, represented by the arrow, of roughly $4/\omega_m$. The $\alpha$-dependence of ${\cal I}(\alpha)$ varies significantly depending on the regime considered. ${\cal I}(\alpha)$ is symmetric about the minimum of zero at $\alpha_g=1/(1+\delta)$ for all $\delta$ and $\eta$. The $\alpha$-dependence tends to be more pronounced further from resonance and for stronger coupling.}\label{mi2}
\end{center}
\end{minipage}
\end{figure}
\begin{figure}[t]
\begin{minipage}{\columnwidth}
\begin{center}
\hspace*{4mm}\includegraphics[scale=0.43]{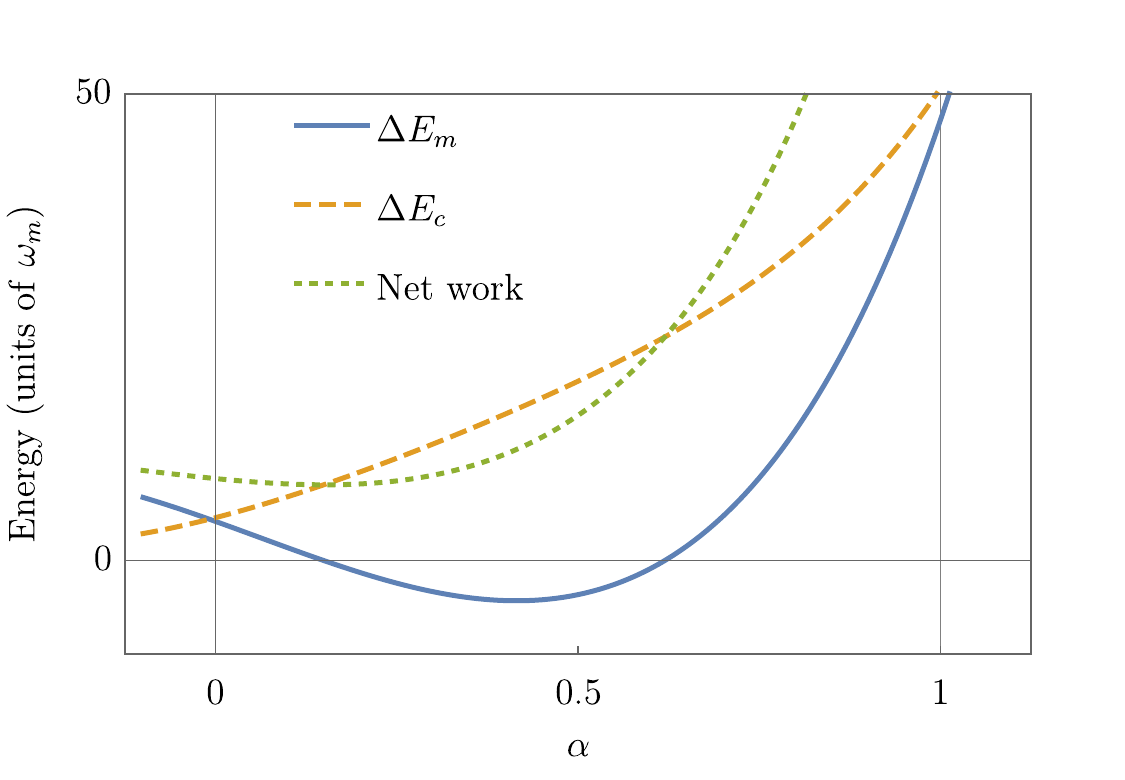}
\caption{$\eta=1$ and $\delta=3$ with $\tau,\,s$, and $\omega_m$ as in Fig. \ref{mi2}. $\beta_c$ corresponds to room temperature while $\beta_m =2\beta_c$. The final subsystem energy changes and net work are plotted with $\alpha$. The net work and $\Delta E_c$ are always positive, while $\Delta E_m$ becomes negative for certain $\alpha$ implying that energy has left the initially cooler system and has entered the initially hotter system. This is due to the non-zero net work input.
}\label{en1}
\end{center}
\end{minipage}
\end{figure}
 
Since the systems are initially uncoupled, it is natural to assume that they are not correlated. Correlations may then build-up due to the subsequent interaction. Figure~\ref{mi2} shows the final mutual information ${\cal I}(\alpha)$ at a time $t$ after the interaction has ceased. It exhibits a diversity of behaviours depending on the values of $\alpha$, $\eta$ and $\delta$. As expected the variations in the mutual information become increasingly pronounced as the coupling strength increases and one moves away from resonance $\delta=1$. As noted in Sec.~\ref{hd}, the weak-coupling resonance regime is gauge-nonrelativistic, i.e., is such that all gauges will produce the same final predictions despite the $H^\alpha(t)$ being non-equivalent. As shown in Figure~\ref{mi2}, if $\alpha=\alpha_g$ the interaction does not generate any correlations for the values of $\eta$ and $\delta$ chosen. Thus, this value of $\alpha$ reproduces the (ostensibly unique) result obtained within the weak-coupling regime even for ultrastrong coupling. The subsystems defined relative to this gauge are those for which the ground state is much closer to the bare vacuum.

To exemplify the importance of our results, we show that due to the time-dependence of the interaction even the {\em qualitative} predictions for energy exchange depend strongly on $\alpha$. To this end we consider a situation where the systems are not initially isolated from their environments. We therefore consider an initial product state of two Gibbs states $\rho(0)=\rho_m^{\rm eq}(\beta_m)\otimes \rho_c^{\rm eq}(\beta_c)$ where $\rho^{\rm eq}_m(\beta_m)= e^{-\beta_x H_x}/{\rm tr}(\cdot),~x=m,\,c$ with $H_m=\omega_m (b^\dagger b+1/2)$ and $H_c=\omega (a^\dagger a+1/2)$. These states result if before their interaction the systems have separately weakly-coupled and equilibrated with Markovian environments at the corresponding temperatures $\beta^{-1}_m$ and $\beta^{-1}_c$. For generality we do not assume these temperatures are equal. If the subsequent light-matter interaction is relatively short on the order of $\omega_m^{-1}$ as in Figs.~\ref{mi2} and \ref{en1}, and is also ultrastrong, then a clear separation of time and energy scales emerges. Weak environmental interactions can therefore be ignored over the time-scales of interest.

Using the unitarity of the dynamics it is straightforward to show that changes in the energies of the subsystems defined by $\Delta E_x = {\rm tr}[\rho_x(t) H_x]-{\rm tr}[\rho_x^{\rm eq} H_x]$ with $x=m,\,c$, are bounded according to $\beta_m\Delta E_m +\beta_c \Delta E_c \geq {\cal I}\geq 0$ \cite{jevtic_maximally_2012,stokes_using_2017}. If the interaction is also such that there is no net input of work, i.e., $\langle \Delta H^\alpha(t)\rangle \equiv \langle \Delta H_0\rangle \equiv \Delta E_m+\Delta E_c=0$, then we obtain $(\beta_m-\beta_c)\Delta E_m\geq 0$. Thus, without a net input of work, energy cannot move counter to the initial temperature gradient. On the other hand if $\Delta E_m + \Delta E_c \neq 0$ then by the end of the interaction the initially cooler system may have lost energy, with an accompanying increase in energy of the initially hotter system. Alternatively, both subsystems may simply gain energy due to the non-zero net work. The final energy that has been exchanged between the systems after the protocol has finished is shown as a function of $\alpha$ in Fig.~\ref{en1}. It is clear that different qualitative thermodynamics can be realised by varying only the parameter $\alpha$, which controls the gauge. We express once more that these qualitative differences in final properties occur even though the subsystems are uniquely defined at both the initial and final times.

\section{Conclusions}\label{conc}

We have studied the implications of gauge-freedom for subsystem properties in QED when dealing with tuneable non-adiabatic, strong-coupling. When the coupling is non-vanishing there are infinitely many non-equivalent definitions of the quantum subsystems. For strong enough coupling ``light" and ``matter" subsystem properties like entanglement and photon number are significantly different for different subsystem definitions. These differences persist in the case of final subsystem predictions found when assuming a time-dependent coupling. The differences become increasingly pronounced as the coupling switching increases in strength and speed.

We have shown directly that the time-dependent coupling assumption can be valid for fast and strong inter- actions, but only if one can identify and choose the correct gauge within which to make the assumption when describing a given physical arrangement. This is necessary in order to obtain even qualitatively reliable pre- dictions. The correct choice of gauge will generally depend on the specific microscopic arrangement being considered. Its determination requires the availability of, and comparison with, a more complete description that explicitly includes the control degrees of freedom. This finding is of major importance for current technological applications including quantum communication, metrology, simulation, and information processing, where the use of time-dependent couplings is widespread and final subsystem properties are of central importance.

{\em Acknowledgment.} This work was supported by the UK Engineering and Physical Sciences Research Council, grant no. EP/N008154/1. We thank Zach Blunden-Codd for useful discussions. 

\bibliography{light_matter.bib}

\newpage

\onecolumngrid
\appendix
\section{Introduction of a different modified current}\label{modc}

\subsection{Charge conservation}

Since the total electric charge is the conserved Noether-charge associated with gauge-symmetry, the non-equivalence of the Lagrangians associated with different gauges can be understood as a consequence of the fact that $\partial_\nu \mu(t)j^\nu= 0$ if and only if ${\dot \mu}=0$. A second implication is that the inhomogeneous Maxwell equations $\nabla \cdot {\bf E}=\mu(t)\rho$ (Gauss' law) and ${\dot {\bf E}}=\nabla \times {\bf B} - \mu(t){\bf J}$ (Ampere's law), cannot be simultaneously satisfied. The method of modelling controllable interactions between given subsystems using a time-dependent coupling parameter is usually adopted at the Hamiltonian level, and is widespread. It is not our intention to advocate such an approach, but merely to understand its implications. The implications above follow from tracing back the conventional approach to the Lagrangian level or to the fundamental equations of motion.

Given the discussion above, one is naturally led to seek a different modified current $\tilde j$, which includes the external control $\mu(t)$, but also satisfies $\partial_\nu {\tilde j}^\nu=0$. Letting ${\tilde j}_0\equiv {\tilde \rho}=\mu\rho$ and then considering ${\dot {\tilde \rho}}$ reveals that the appropriate modified three-current ${\tilde {\bf J}}$ must satisfy
\begin{align}\label{mod0}
\nabla \cdot {\tilde {\bf J}} = \mu(t) \nabla \cdot {\bf J} -{\dot \mu}(t)\rho. 
\end{align}
This clearly necessitates the addition of a non-trivial term to the naive modified current $\mu(t){\bf J}$. Solving Eq.~(\ref{mod0}) for ${\tilde {\bf J}}$ requires inverting the divergence operator, which introduces a new arbitrary element into the formalism. The solution can be expressed as
\begin{align}
{\tilde {\bf J}} = \mu(t){\bf J} + {\dot \mu}(t){\bf P} 
\end{align}
where
\begin{align}\label{Pol}
{\bf P}({\bf x}) = -\int d^3 x' {\bf g}({\bf x},{\bf x}')\rho({\bf x}')
\end{align}
in which $\nabla\cdot {\bf g}({\bf x},{\bf x}')=\delta({\bf x}-{\bf x}')$. The polarisation ${\bf P}$ satisfies $-\nabla \cdot {\bf P} = \rho$ identically, but has arbitrary transverse component, because $\nabla \cdot {\bf g}_{\rm T}({\bf x},{\bf x}')\equiv 0$. Defining ${\tilde {\bf P}}=\mu(t){\bf P}$ one recovers the well-known continuity and polarisation relations in terms of the modified quantities;
\begin{align}
\partial_\nu {\tilde j}^\nu =0,~~-\nabla\cdot {\tilde {\bf P}} ={\tilde \rho}.
\end{align}
The complete construction of the charge and current densities using auxiliary fields requires the introduction of the magnetisation ${\bf M}$ such that ${\bf J}={\dot {\bf P}}+\nabla \times {\bf M}$. The modified magnetisation required to give ${\tilde {\bf J}}={\dot {\tilde {\bf P}}}+\nabla\times {\tilde {\bf M}}$ is therefore simply ${\tilde {\bf M}}=\mu(t){\bf M}$. Since ${\bf P}$ and ${\bf M}$ are auxiliary fields for $\rho$ and ${\bf J}$, they can be viewed as material analogs of the electromagnetic potentials, which are auxiliary fields for ${\bf E}$ and ${\bf B}$.

Like the polarisation ${\bf P}$ the definition of the magnetisation ${\bf M}$ also possesses an arbitrary freedom. Indeed,  the definitions of these auxiliary quantities in terms of the charge and current densities possess the same structure as the inhomogeneous Maxwell equations, but these equations are not supplemented by any homogeneous Maxwell-type equations. It follows that any transformation of ${\bf P}$ and ${\bf M}$ that leaves the defining inhomogeneous equations invariant is permissible. Thus, $j$ is invariant under a transformation by pseudo-magnetic and pseudo-electric fields as
\begin{align}\label{pmt}
{\bf P}\to {\bf P}+\nabla \times {\bf U},~~{\bf M} \to {\bf M}-\nabla U_0 -{\dot {\bf U}}
\end{align}
where $U$ is an arbitrary pseudo-four-potential. The fields are in turn invariant under a gauge-transformation $U_\mu\to U_\mu -\partial_\mu \chi$ where $\chi$ is arbitrary. The modified current ${\tilde {\bf J}}$ is not invariant under the transformation
 (\ref{pmt}), which results in ${\tilde {\bf J}} \to {\tilde {\bf J}}+{\dot \mu}(t)\nabla\times {\bf U}$.
 
If we now replace the naive modified current $\mu(t){\bf J}$ that appears in Eq.~(2) of the main text with ${\tilde {\bf J}}$ we obtain
\begin{align}\label{Ltil}
{\tilde L} =L + {\dot \mu}(t)\int d^3 x\,  {\bf P}\cdot {\bf A}.
\end{align}
Since ${\tilde L}$ is not equivalent to $L$, it does not possess the same properties under a gauge transformation of the electromagnetic potentials, which gives
\begin{align}
{\tilde L} \to {\tilde L} + {d\over dt} \int d^3 x {\tilde \rho}\chi
\end{align}
as desired. However, it can be seen immediately from Eq.~(\ref{Ltil}) that unlike the original Lagrangian $L$, under the transformation (\ref{pmt}) the Lagrangian ${\tilde L}$ transforms to an equivalent Lagrangian if and only if ${\dot \mu}=0$. Thus, our construction of ${\tilde j}$, ${\tilde {\bf P}}$ and ${\tilde L}$ has replaced one gauge non-invariance with another. The inhomogeneous Maxwell equations are now simultaneously satisfied when written in terms of the modified quantities, but the modified current ${\tilde {\bf J}}$ is not invariant under the transformation (\ref{pmt}) and therefore neither is Ampere's law when written in terms of ${\tilde {\bf J}}$. We stress that the freedom to choose the transverse component of ${\bf P}$ is an important freedom within the theory, and is no less significant than the freedom to choose the potentials. Indeed, as we will see in Sec.~\ref{gfix} the freedom to choose ${\bf P}_{\rm T}$ is what gives rise to the well-known Poincar\'e-gauge dipolar interaction Hamiltonian $-e{\bf r}\cdot {\bf \Pi}({\bf 0})$.

Essentially the same result as Eq.~(\ref{Ltil}) above can be obtained if instead of considering the current, one considers the interaction Lagrangian. The standard interaction Lagrangian density $j^\mu A_\mu$ is not gauge-invariant, rather, under a gauge transformation $A_\mu\to A_\mu -\partial_\mu\chi$ it changes as
\begin{align}\label{Ligt}
{\cal L}_I =-\int d^3 x j^\mu A_\mu \longrightarrow  -\int d^3 x j^\mu A_\mu + {d\over dt} \int d^3x \rho \chi.
\end{align}
Since the remaining Lagrangian components are manifestly gauge-invariant, according to Eq.~(\ref{Ligt}) the total Lagrangian changes under a gauge transformation by the addition of a total time derivative meaning that the result is equivalent, but not identical. This equivalence no longer holds if the interaction Lagrangian is $L_I(t) = \mu(t){\cal L}_I$. The additional term that results from the gauge transformation is now $\mu(t){d\over dt}\int d^3x\, \rho\chi$, which is not a total time derivative. However, it is clear that this non-equivalence could be avoided if one starts with a manifestly gauge-invariant interaction Lagrangian from the outset. Such an interaction Lagrangian is given by
\begin{align}
{\cal L}'_I =\int d^3 x \left[{\bf P}\cdot {\bf E}+{\bf M}\cdot {\bf B}\right] = -\int d^3 x \left[j^\mu A_\mu +{d\over dt}{\bf P}\cdot {\bf A}\right],
\end{align}
which is clearly invariant under a gauge transformation. However, under the transformation (\ref{pmt}) ${\cal L}'_I$ transforms to an equivalent but different form as
\begin{align}\label{lpmt}
{\cal L}'_I \to {\cal L}'_I -{d\over dt}\int d^3 x {\bf B}\cdot {\bf U}
\end{align}
where we have used Faraday's law ${\dot {\bf B}}=-\nabla \times {\bf E}$. The original interaction  $j^\mu A_\mu$ involves the physical matter fields $j^\mu$ and the auxiliary electromagnetic fields $A_\mu$, while the interaction ${\cal L}'_I$ involves the physical electromagnetic fields and the auxiliary matter fields. Thus, if we define a new time-dependent interaction Lagrangian by
\begin{align}\label{lagint2}
L'_I(t) =\mu(t){\cal L}'_I=-\mu(t) \int d^3 x \left[j^\mu A_\mu +{d\over dt}{\bf P}\cdot {\bf A}\right] = \mu(t) \int d^3 x \left[{\bf P}\cdot {\bf E}+{\bf M}\cdot {\bf B}\right],
\end{align}
then we obtain a total Lagrangian that despite including the external control $\mu(t)$, is invariant under a gauge-transformation of the potentials. However, as is to be expected on the basis of the transformation property (\ref{lpmt}), this comes at the price of producing a non-equivalent Lagrangian under the transformation (\ref{pmt}). Indeed, the interaction Lagrangian in Eq.~(\ref{lagint2}) is actually what results from using the modified quantities ${\tilde j}$ and ${\tilde {\bf P}}$ within ${\cal L}'_I$. To show this we denote by ${\tilde L}'$ the total Lagrangian obtained by replacing in $L'$, the current $j$ and polarisation ${\bf P}$ with their modified counterparts, and note that a quick calculation gives
\begin{align}\label{lagpt}
{\tilde L}'\equiv \, &L_{\rm m} +L_{\rm TEM}+{\mu(t)^2 \over 2}\int d^3 x \, \rho \phi_{\rm Coul} -\int d^3 x\left[ {\tilde j}^\mu A_\mu +{d\over dt}{\tilde {\bf P}}\cdot {\bf A}\right]\nonumber \\ = \,& L_{\rm m} +L_{\rm TEM}+{\mu(t)^2 \over 2}\int d^3 x \, \rho \phi_{\rm Coul} -\mu(t)\int d^3 x\left[ j^\mu A_\mu +{d\over dt}{\bf P}\cdot {\bf A}\right].
\end{align}
It is now readily verified that
\begin{align}
{\tilde L}' =  {\tilde L} - {d\over dt}\int d^3 x \,{\tilde {\bf P}}\cdot {\bf A} =  L - \mu(t){d\over dt}\int d^3 x \,{\bf P}\cdot {\bf A} 
\end{align}
where ${\tilde L}$ is given by Eq.~(\ref{Ltil}). The new Lagrangian ${\tilde L}'$ is equivalent to ${\tilde L}$ which was the result we obtained by replacing $j$ and ${\bf P}$ with ${\tilde j}$ and ${\tilde {\bf P}}$ in $L$. The only difference between ${\tilde L}$ and ${\tilde L}'$ is that under a gauge transformation (\ref{gauge}), ${\tilde L}$ transforms into an equivalent form, whereas ${\tilde L}'$ is invariant. Whichever of these equivalent Lagrangians is considered, it is clear that unlike the original Lagrangian $L$, neither provides an equivalent Lagrangian under the transformation (\ref{pmt}). Conversely $L$ is invariant under the transformation (\ref{pmt}), but does not provide an equivalent Lagrangian under a gauge-transformation of the electromagnetic potentials. 

\subsection{Gauge-fixing}\label{gfix}

In conventional approaches to non-relativistic QED all gauge-redundancies are eliminated simultaneously through a gauge-fixing constraint that has the form
\begin{align}\label{C}
\int d^3 x' {\bf g}({\bf x}',{\bf x})\cdot {\bf A}({\bf x}')=0
\end{align}
where ${\bf g}$ must be the {\em same} choice of green's function as is made in Eq. (\ref{Pol}). Choosing the potentials
\begin{align}
{\bf A}={\bf A}_{\rm T}+\nabla \chi,~~A_0=\mu(t)\phi_{\rm Coul}-\partial_t\chi
\end{align}
where
\begin{align}
\chi({\bf x})=\int d^3 x' {\bf g}_{\rm T}({\bf x}',{\bf x})\cdot {\bf A}_{\rm T}({\bf x}')
\end{align}
means that Eq. (\ref{C}) is satisfied identically. The freedom to choose a gauge now reduces to the freedom to choose ${\bf g}_{\rm T}$, which uniquely specifies both the four-potential $A$ and the polarisation ${\bf P}$. Two special cases are given by the Coulomb gauge ${\bf g}_{\rm T}=0$ and the Poincare-gauge ${\bf g}_{{\rm T},i}({\bf x},{\bf x}')= -x'_j\int_0^1 d\lambda \, \delta_{ij}^{\rm T}({\bf x}-\lambda{\bf x'})$.

The invariance of ${\cal L}'$ and of ${\tilde L}'$ under gauge transformations requires that ${\bf P}$ is not altered by the gauge-transformation, but this is no longer the case if both $A$ and ${\bf P}$ are simultaneously determined by ${\bf g}_{\rm T}$. A new choice of ${\bf g}_{\rm T}$ via ${\bf g}_{\rm T}\to {\bf g}_{\rm T}'$ will result in a gauge transformation of both $A$ {\em and} ${\bf P}$. The latter will transform as in (\ref{pmt}) with
\begin{align}
\nabla \times {\bf U}({\bf x}) = \int d^3 x' \left[{\bf g}'_{\rm T}({\bf x},{\bf x}')-{\bf g}_{\rm T}({\bf x},{\bf x}')\right] \rho({\bf x'}).
\end{align}
This (gauge) freedom in ${\bf P}_{\rm T}$ is central to quantum optics and molecular electrodynamics, because when transforming from the Coulomb to Poincar\'e (multipolar) gauges the additional contribution $\nabla\times {\bf U}={\bf P}_{\rm T}^1$ provides the dominant interaction Hamiltonian $\int d^3x\, [{\bf \Pi}\cdot {\bf P}_{\rm T}^1+({\bf P}_{\rm T}^1)^2]$. This is the only non-vanishing interaction term in the dipole approximation, wherein the contribution $\int d^3x\, {\bf \Pi}\cdot {\bf P}_{\rm T}^1$ possesses the well-known form $-e{\bf r}\cdot {\bf \Pi}({\bf 0})$.

If ${\dot \mu}=0$ then all forms of the Lagrangian are equivalent. However, if ${\dot \mu}\neq 0$ we have only been able to construct Lagrangians that are at best partially invariant under a {\em complete} gauge-transformation of both electromagnetic and material potentials. If, in particular, we impose the constraint (\ref{C}), which all standard non-relativistic gauges satisfy, then $\int d^3x \,{\bf P}\cdot {\bf A}=0$, implying that $L$, ${\tilde L}$ and ${\tilde L}'$ all coincide. Moving afterwards to the canonical formalism results in non-equivalent Hamiltonians as was shown in Sec.~II C of the main text.

\section{Full description of the dipole's motion}\label{fquant}

We consider a two-charge system comprised of a charge $e_1=-e$ with mass $m_1$ at ${\bf r}_1$ and a charge $e_2=e$ with mass $m_2$ at ${\bf r}_2$. We introduce relative and centre-of-mass coordinates as
\begin{align}
{\bf r}={\bf r}_1-{\bf r}_2,\qquad \qquad
{\bf R} = {m_1{\bf r}_1 + m_2{\bf r}_2 \over M}
\end{align}
where $M=m_1+m_2$. We start with the standard Lagrangian
\begin{align}
L=&{1\over 2}m_1{\dot {\bf r}}_1 + {1\over 2}m_2{\dot {\bf r}}_2 - \int d^3 x\left[j^\mu A_\mu + {1\over 4}F^{\mu\nu}F_{\mu\nu}\right] \nonumber \\ =& {1\over 2}m_1{\dot {\bf r}}_1 + {1\over 2}m_2{\dot {\bf r}}_2 - V({\bf r}_1-{\bf r}_2)+ \int d^3 x \left[\rho\partial_t\chi^\alpha +{\bf J}\cdot {\bf A} +{1\over 2}({\bf E}_{\rm T}^2-{\bf B}^2)\right] \nonumber \\ =&
{1\over 2}m{\dot {\bf r}} + {1\over 2}M{\dot {\bf R}} - V({\bf r})+ \int d^3 x \left[{\bf J}\cdot {\bf A}_{\rm T} -{d\over dt}{\bf P}^\alpha_{\rm T}\cdot {\bf A}_{\rm T} +{1\over 2}({\bf E}_{\rm T}^2-{\bf B}^2)\right] 
\end{align}
where $m=m_1m_2/M$ and $V({\bf r}_1-{\bf r}_2)=V({\bf r})$ is the inter-charge Coulomb energy. The infinite Coulomb self-energies have been ignored. The remaining quantities are given by
\begin{align}
\rho({\bf x})&=e[\delta({\bf x}-{\bf r}_2)-\delta({\bf x}-{\bf r}_1)], \label{s1}\\
{\bf J}({\bf x})&=e{\dot {\bf r}}_2\delta({\bf x}-{\bf r}_2)-e{\dot {\bf r}}_1\delta({\bf x}-{\bf r}_1), \\
{\bf P}^\alpha_{\rm T}({\bf x}) &= -\int d^3 {\bf g}^\alpha_{\rm T}({\bf x},{\bf x}')\rho({\bf x}'),\\
{\bf A}&= {\bf A}_{\rm T}+\nabla\chi^\alpha \label{s4}
\end{align}
where
\begin{align}
g^\alpha_{{\rm T},i}({\bf x},{\bf x}') &= -\alpha({\bf x}'-{\bf R})_j \int_0^1 d\lambda \,\delta_{ij}^{\rm T}({\bf x}-{\bf R}-\lambda({\bf x}'-{\bf R})),\\
\chi^\alpha({\bf x}) &= \int d^3 x' {\bf g}^\alpha_{\rm T}({\bf x}',{\bf x})\cdot {\bf A}_{\rm T}({\bf x}').
\end{align}
Here ${\bf g}_{\rm T}^\alpha$ is chosen such that ${\bf g}^1_{\rm T}$ gives the usual multipolar transverse polarisation. Notice however that this means that ${\bf g}_{\rm T}^\alpha$ depends on the centre-of-mass position ${\bf R}$. 

The electric dipole approximation (EDA) is obtained by retaining only the leading order term in the multipole expansion of $\rho$ about ${\bf R}$, which for ${\dot {\bf R}}\neq {\bf 0}$ results in
\begin{align}
\rho({\bf x}) &= e{\bf r}\cdot \nabla\delta({\bf x}-{\bf R}),\\
{\bf J}({\bf x}) &= -e{\dot {\bf r}}\delta({\bf x}-{\bf R})+e{\dot {\bf R}}({\bf r}\cdot \nabla)\delta({\bf x}-{\bf R}).\label{J2}\\
P^1_{{\rm T},i}({\bf x})&=-er_j\delta_{ij}^{\rm T}({\bf x}-{\bf R}).
\end{align}
The second term in Eq. (\ref{J2}) vanishes if and only if the atom is at rest in the lab frame. This term is vital for ensuring that the correct R\"ontgen interaction due to atomic motion is included. Substituting these expressions into Eqs. (\ref{s1})-(\ref{s4}) and using the resulting expressions in the Lagrangian gives the Lagrangian within the EDA. This can be taken as the starting point for the canonical formalism with ${\bf r}$, ${\bf R}$ and ${\bf A}_{\rm T}$ as canonical coordinates. The momenta conjugate are denoted ${\bf p}$, ${\bf K}$ and ${\bf \Pi}$ respectively. The resulting Hamiltonian is
\begin{align}\label{ham}
H^\alpha&= {1\over 2M}[{\bf K} +e({\bf r}\cdot \nabla){\bf A}_{\rm T}({\bf R})-e\alpha \nabla_{\bf R}{\bf r}\cdot {\bf A}_{\rm T}({\bf R})]^2 + {1\over 2m}[{\bf p}+e(1-\alpha){\bf A}_{\rm T}({\bf R})]^2 + V({\bf r})+  {1\over 2}\int d^3x \left[({\bf \Pi}+{\bf P}^\alpha_{\rm T})^2+{\bf B}^2\right] \nonumber \\ &= R_{\alpha 0}H^0R_{0\alpha}
\end{align}
where
\begin{align}
P_{{\rm T},i}^1({\bf x}) = -er_j \delta_{ij}^{\rm T}({\bf x}-{\bf R}),\qquad R_{0\alpha} = \exp\left(-i\alpha\int d^3 x{\bf P}^1_{\rm T}\cdot {\bf A}_{\rm T}\right)=e^{i\alpha e {\bf r}\cdot {\bf A}_{\rm T}({\bf R})}.
\end{align}
At this stage the theory is completely gauge-invariant because the $H^\alpha$ are unitarily equivalent. The predictions for any physical observable are independent of the choice of gauge $\alpha$. It is nevertheless the case that the quantum subsystems are defined differently in each different gauge. Subsystem properties like photon number and entanglement will generally depend on the definitions chosen, that is, they will depend on the choice of gauge relative to which the subsystems are defined.

\subsection{Approximation of externally prescribed uniform gross motion in the Hamiltonian}

The approximation of an externally controlled coupling between the dipole and the field results from the assumption that the dynamical variable ${\bf R}(t)={\dot {\bf R}}t$ (up to a constant initial position) is external and prescribed. This means that ${\dot {\bf R}}$ is also prescribed. With this assumption the Hamiltonian in Eq. (\ref{ham}) becomes that of a bipartite quantum system, and reads as
\begin{align}\label{ham2d}
H^\alpha(t) = & {1\over 2}M{\dot {\bf R}}^2  + {1\over 2m}[{\bf p}+e(1-\alpha){\bf A}_{\rm T}({\bf R})]^2 + V({\bf r}) +  {1\over 2}\int d^3x \left[({\bf \Pi}+{\bf P}^\alpha_{\rm T})^2+{\bf B}^2\right],
\end{align}
where now ${\bf R}$ and ${\dot {\bf R}}$ are given classical variables. Since the first kinetic term $M{\dot {\bf R}}^2/2$ is not operator-valued and for uniform motion is also constant in time, it can be ignored. Before approximating ${\bf R}(t)$ as external, the Hamiltonians in Eq. (\ref{ham}) were seen to be equivalent, but the $H^{\alpha}(t)$ in Eq. (\ref{ham2d}) are not equivalent for different $\alpha$, being related by
\begin{align}\label{hrel2}
H^{\alpha'}(t) = R_{\alpha\alpha'}(t)H^\alpha(t)R_{\alpha'\alpha}(t)
\end{align}
where $R_{\alpha\alpha'}(t)=\exp[-ie(\alpha-\alpha'){\bf r}\cdot {\bf A}_{\rm T}({\bf R}(t))]$. Equation~(\ref{hrel2}) shows that Hamiltonians associated with different gauges are not equivalent, because 
\begin{align}
H^{\alpha'}(t) \neq R_{\alpha\alpha'}(t)H^\alpha(t)R_{\alpha' \alpha}(t) + i{\dot R}_{\alpha\alpha'}(t)R_{\alpha'\alpha}(t)
\end{align}
where the right-hand-side of this inequality is equivalent to $H^\alpha(t)$. To obtain the Hamiltonian in Eq. (\ref{hamalph}) of the main text, which was obtained by assuming a time-dependent coupling $\mu(t)$, one requires only that ${\bf A}_{\rm T}({\bf R}(t))$ can be written ${\bf A}_{\rm T}({\bf R}(t)) = \mu(t){\bf A}_{\rm T}({\bf 0})$. This is indeed the case in the example we consider in the main text and in Appendix \ref{ma} whereby an atom moves in and out of a Gaussian cavity beam for which ${\bf A}_{\rm T}({\bf x})$ has the form ${\bm \varepsilon}A({\bf x})$, and we also assume that ${\bf r}=r{\bm \varepsilon}$.

\subsection{Approximation of externally prescribed uniform gross motion in the Lagrangian}

If we approximate ${\bf R}={\dot {\bf R}}t$ as external at the Lagrangian level then the remaining variables are ${\bf r}$ and ${\bf A}_{\rm T}$. The $\alpha$-gauge Hamiltonian is
\begin{align}
{\tilde H}^\alpha =& {\bf p}\cdot {\dot {\bf r}}+\int d^3x \, {\bf \Pi}\cdot {\bf A}_{\rm T} - L \nonumber \\
=& {1\over 2m}[{\bf p}+e(1-\alpha){\bf A}_{\rm T}({\bf R})]^2 + V({\bf r}) +  {1\over 2}\int d^3x \left[({\bf \Pi}+{\bf P}^\alpha_{\rm T})^2+{\bf B}^2\right] + e{\dot {\bf R}}\cdot \left[({\bf r}\cdot \nabla){\bf A}_{\rm T}({\bf R})\right]-e\alpha ({\dot {\bf R}}\cdot \nabla){\bf r}\cdot {\bf A}_{\rm T}({\bf R})
\end{align}
where the constant kinetic energy $M{\dot {\bf R}}^2/2$, which depends only on the external control, has been neglected.
This is the Hamiltonian given in Eq. (\ref{h45}) of the main text. As Schr\"odinger picture operators these Hamiltonians are related by 
\begin{align}
{\tilde H}^{\alpha'}(t) = R_{\alpha\alpha'}(t){\tilde H}^\alpha(t)R_{\alpha'\alpha}(t) +i{\dot R}_{\alpha\alpha'}(t)R_{\alpha'\alpha}(t)
\end{align}
where, as before, $R_{\alpha\alpha'}(t)=\exp[ie(\alpha-\alpha'){\bf r}\cdot {\bf A}_{\rm T}({\bf R}(t))]$.

\section{Atom moving in and out of a Fabry-Perot cavity}\label{ma}

In this appendix we specialise the Hamiltonian derived above in Eq. (\ref{ham2d}) to describe the interaction between a Fabry-Perot cavity and an oscillating dipole at an arbitrary position within the cavity.

\subsection{Quantisation of the free cavity}\label{fp}

We consider a Fabry-Perot cavity consisting of parallel mirrors in the $xy$-plane separated by a distance $L$. In the $z$-direction the electromagnetic field satisfies periodic boundary conditions, with a Gaussian profile in the perpendicular direction $x{\bf \hat{x}}+y{\bf {\hat y}}$ \cite{kogelnik_laser_1966}. We restrict our attention to the fundamental Gaussian mode in the perpendicular direction. Although not necessary, for simplicity we also consider only the fundamental standing wave mode in the $z$-direction. It is straightforward to extend this model to the multi-mode case that includes more standing-wave modes in the $z$-direction. One could also consider additional Gauss-Hermite or Gauss-Laguerre modes in the perpendicular direction.

In the present case the single cavity mode is described by a pure Gaussian beam propagating in the $z$-direction such that classically the transverse vector potential is
\begin{align}\label{gausA}
{\bf A}(t,{\bf x}) = {\bm \varepsilon} \,{\cal A} a\, u({\bf x})e^{-i\omega t}+{\rm c.c.}
\end{align}
where ${\bm \varepsilon}$ is a transverse polarisation in the $xy$-plane and $u({\bf x})e^{-i\omega t}$ satisfies the paraxial scalar wave equation \cite{kogelnik_laser_1966,wunsche_quantization_2004,zangwill_modern_2012}. Anticipating the transition to the quantum theory we have written the space and time-independent amplitude ${\cal A}a$ as the product of a real normalisation ${\cal A}$ and a complex number $a$. We have also neglected a small non-transverse component in the $z$-direction \cite{wunsche_quantization_2004,zangwill_modern_2012}. We define ${\bf \Pi}(t,{\bf x}) = {\dot {\bf A}}(t,{\bf x})\equiv -{\bf E}_{\rm T}(t,{\bf x})$ such that the cavity energy is
\begin{align}\label{caven}
H_l &= {1\over 2}\int' d^3 x \,[{\bf E}_{\rm T}({\bf x})^2+ {\bf B}({\bf x})^2]  =\int' d^3 x \,{\bf \Pi}({\bf x})^2
\end{align}
where $\int'$ indicates that spatial integration is restricted to the cavity length $L$ in the $z$-direction, and ${\bf B}=\nabla \times {\bf A}$. We have assumed that the magnetic and electric energies are the same in the free theory.

To obtain an explicit expression for $H_l$ that can be quantised, we consider the fundamental Gaussian mode solution to the paraxial wave equation $u({\bf x})e^{-i\omega t}$ such that \cite{zangwill_modern_2012}
\begin{align}\label{u}
u({\bf x}) = {w_c\over w(z)} e^{-(x^2+y^2)/w(z)^2} e^{ik(x^2+y^2)/2R(z)+i\theta(z) + ikz}
\end{align}
with $(0,0,k)$ the wave-vector such that $k=\omega$ and
\begin{align}
z_R &= {1\over 2}kw_c^2,\qquad 
\,w(z) = w_c\sqrt{1+\left({z\over z_R}^2\right)},\nonumber \\
R(z) &=z+{z_R^2\over z}, \qquad 
\theta(z) = -\arctan{z\over z_R},
\end{align}
where $w_c$ denotes the beam waist. For $L\ll z_R$ we have $w(z)\approx w_c$, $k(x^2+y^2)/2R(z)\approx 0$ and $\theta(z)\approx -\pi/2$. In this limit Eq.~(\ref{u}), reduces to
\begin{align}\label{phi}
u({\bf x})\approx \phi({\bf x}) = e^{ikz}e^{-(x^2+y^2)/w_c^2}
\end{align}
where we have ignored a global phase $e^{-i\pi/2}$. We define the cavity volume by
\begin{align}\label{vol2}
v ={1\over 2} \int' d^3 x |\phi({\bf x})|^2 = {\pi w_c^2 L \over 2}
\end{align}
and choose the normalisation ${\cal A}=1/\sqrt{2\omega v}$, such that substitution of Eq.~(\ref{gausA}) into the right-hand-side of Eq.~(\ref{caven}) yields
\begin{align}\label{caven2}
H_l ={\omega\over 2}(a^* a + a a^*)={v\over 2}({\bf \Pi}^2+\omega^2{\bf A}^2)
\end{align}
where ${\bf A}\equiv {\bf A}({\bf x}={\bf 0})$ and ${\bf \Pi}\equiv {\bf \Pi}({\bf x}={\bf 0})$. This cavity Hamiltonian is formally identical to the bare-cavity Hamiltonian of Sec.~\ref{hd}, and in the free (non-interacting) theory it is $\alpha$-independent. In obtaining Eq.~(\ref{caven2}) we have used
\begin{align}\label{int1}
\int' d^3 x\, \phi({\bf x})^2 = \int_l^{l+L} dz \,e^{2i kz} \int dx dy\, e^{-(x^2+y^2)/w_c^2} =0,
\end{align}
where $l$ is arbitrary such that $l$ and $l+L$ are the positions of the two cavity mirrors along the $z$-axis. Equation~(\ref{int1}) follows from the vanishing of the $z$-integral due to the periodic boundary conditions in the $z$-direction; $k=n\pi /L,~n=0,1,2,3...~$.

Quantisation is now straightforward via the replacement of the complex numbers $a$ and $a^*$ with bosonic operators $a$ and $a^\dagger$ such that $[a,a^\dagger]=1$. We thereby obtain the mode expansions
\begin{align}
{\bf A}(t,{\bf x}) &=  {{\bm \varepsilon}\over \sqrt{2\omega v}} [\phi^*({\bf x}) a^\dagger (t) +\phi({\bf x}) a(t)], \label{mod1} \\
{\bf \Pi}(t,{\bf x}) &= i{\bm \varepsilon} \sqrt{\omega \over 2v} [\phi^*({\bf x}) a^\dagger (t) -\phi({\bf x}) a(t)],\label{mod2}
\end{align}
where $a(t)=a e^{-i\omega t}$ in the free theory. All non-zero equal-time canonical commutation relations are obtained from Eqs.~(\ref{mod1}) and (\ref{mod2}) using $[a,a^\dagger]=1$;
\begin{align}
[A_i(t,{\bf x}),\Pi_j(t,{\bf x}')]&=i {\varepsilon_i \varepsilon_j \over 2v}[\phi({\bf x})\phi^*({\bf x}') + \phi^*({\bf x})\phi({\bf x}')],\label{coms1} \\
[A_i(t,{\bf x}),A_j(t,{\bf x}')] &= {\varepsilon_i \varepsilon_j \over 2\omega v}[\phi({\bf x})\phi^*({\bf x}') - \phi^*({\bf x})\phi({\bf x}')],\label{coms2} \\
[\Pi_i(t,{\bf x}),\Pi_j(t,{\bf x}')] &= \omega^2 [A_i(t,{\bf x}),A_j(t,{\bf x}')] \label{coms3}.
\end{align}
In particular we have $[A_i,\Pi_j]=i \varepsilon_i \varepsilon_j/v$ and $[A_i,A_j]=0=[\Pi_i,\Pi_j]$ in agreement with Sec.~III of the main text.

The violation of relativistic causality implied by the non-vanishing commutators of fields at spacelike separated events is a result of the approximations made, namely the restriction to a single radiation mode and the paraxial approximation. The single-mode approximation eliminates the spatio-temporal structure necessary to elicit causality and has been discussed in this context recently in Ref.~\cite{munoz_resolution_2018}. These authors consider the propagation direction only and show that by including more standing wave modes consistency with relativistic causality is recovered. Here, our aim is to study the role of the gauge-parameter $\alpha$ in the light-matter interaction and for this purpose it suffices to restrict attention to the fundamental mode. As noted at the beginning of this section the single-mode approximation is certainly not necessary and has been used here for simplicity. Without requiring any essentially new theoretical machinery one can extend the present treatment in a straightforward manner to include more modes in the transverse direction or in the $z$-direction. Within the single-mode treatment, which is adequate for the present purpose, the canonical commutation relations (\ref{coms1})-(\ref{coms3}) are necessary for the formal self-consistency of the framework developed.

\subsection{Cavity-dipole interaction}

Using the above expressions for the field of a Gaussian cavity mode the full Hamiltonian for atomic motion in and out of the cavity is given by Eq. (\ref{ham}). To preserve gauge-invariance we must also perform the single-mode approximation within the material polarisation. The appropriate approximation can be deduced by inspection of the linear polarisation interaction component, which expressed in ${\bf k}$-space reads
\begin{align}\label{polint}
\int d^3 k \, {\bf P}_{\rm T}^\dagger({\bf k}) \cdot {\bf \Pi}_{\rm T}({\bf k})= i\int d^3 k \sqrt{\omega\over 2}\sum_\lambda {\bm \varepsilon}_\lambda\cdot  \left[{\bf P}_{\rm T}({\bf k})a^\dagger_\lambda({\bf k})-{\bf P}_{\rm T}^\dagger({\bf k}) a_\lambda({\bf k})\right]
\end{align}
where $a_\lambda({\bf k})$ is the annihilation operator for a photon with momentum ${\bf k}$ and polarisation $\lambda$, and ${\bm \varepsilon}_\lambda$ is the corresponding polarisation unit vector, which is orthogonal to ${\bf k}$. In writing Eq.~(\ref{polint}) we have used the Hermiticity of the real-space fields. Discretising the modes in a volume $v$ with periodic boundary conditions and restricting to a single mode ${\bf k}\lambda$, the interaction becomes 
\begin{align}
\int d^3 k \, {\bf P}_{\rm T}^\dagger({\bf k}) \cdot {\bf \Pi}_{\rm T}({\bf k})\longrightarrow  i\sqrt{v\omega\over 2}{\bm \varepsilon} \cdot  \left[{\bf P}_{\rm T}({\bf k})a^\dagger-{\bf P}_{\rm T}^\dagger({\bf k}) a\right].
\end{align}
For the Gaussian cavity this must be equal to $-e{\bf r}\cdot {\bf \Pi}({\bf R})$ where ${\bf \Pi}({\bf R})$ is given by Eq. (\ref{mod2}). It follows that the single-mode approximation of ${\bf P}_{\rm T}({\bf k})$ appropriate for the Gaussian cavity is
\begin{align}
{\bm \varepsilon}\cdot {\bf P}_{\rm T}^\alpha({\bf k}) = -{e\over v} ( {\bm \varepsilon}\cdot {\bf r})\phi^*({\bf R}).
\end{align}
The polarisation self energy term in the Hamiltonian is therefore
\begin{align}
&{1\over 2}\int d^3x {\bf P}_{\rm T}({\bf x})^2 = {1\over 2}\int d^3k |{\bf P}_{\rm T}({\bf k})|^2 = {1\over 2}\int d^3k \sum_\lambda ({\bm \varepsilon}_\lambda \cdot {\bf P}_{\rm T}({\bf k}))({\bm \varepsilon}_\lambda \cdot {\bf P}_{\rm T}({\bf k}))^*\nonumber \\ &\longrightarrow {v\over 2}({\bm \varepsilon}\cdot {\bf P}_{\rm T}({\bf k}))({\bm \varepsilon} \cdot {\bf P}_{\rm T}({\bf k}))^* = {e^2 \over 2v} r^2 |\phi({\bf R})|^2
\end{align}
where $r={\bm \varepsilon}\cdot {\bf r}$. We can now specify all terms within the complete Hamiltonian in Eq.~(\ref{ham}) for the case of a single-mode Gaussian cavity. The Hamiltonians of different gauges are unitarily related and therefore equivalent. Assuming for simplicity that ${\bf r}=r{\bm \varepsilon}$ and ${\bf p}=p{\bm \varepsilon}$, when we approximate ${\bf R}(t)$ as external we obtain Eq.~(31) given in the main text. The resulting Hamiltonians continue to be unitarily related, but are no longer equivalent. Thus, it is the approximation of treating ${\bf R}(t)$ as external that results in non-equivalence between gauges.

\subsection{Mutual information}

Without loss of generality we can consider cavity mirrors located at $z=\pm L/2$ centred at $(0,0)$ in the $xy$-plane. Any prescribed dipolar motion may now be considered. The simplest case consists of uniform motion ${\dot {\bf R}}=-\nu {\hat {\bf x}}$ starting from rest at the point $h{\hat {\bf x}}$, which yields the path ${\bf R}(t)={\hat {\bf x}}(h-\nu t)$. 
Quite generally paths satisfying ${\hat {\bf z}}\cdot {\bf R}(t)=0$ for all $t$ have the property that the Hamiltonian in Eq.~(31) of the main text with ${\bf R}={\bf R}(t)$ is identical to that in Eq.~(23) of the main text if the time-dependent coupling function therein is taken as $\mu(t)= \phi({\bf R}(t))$. For uniform motion the coupling function is $\mu(t)=e^{-(h-\nu t)^2/w_c^2}$ in the case of uniform motion. 
\\ \\

\begin{figure}[H]
\begin{minipage}{\columnwidth}
\begin{center}
\hspace*{-0.2cm}\includegraphics[scale=0.44]{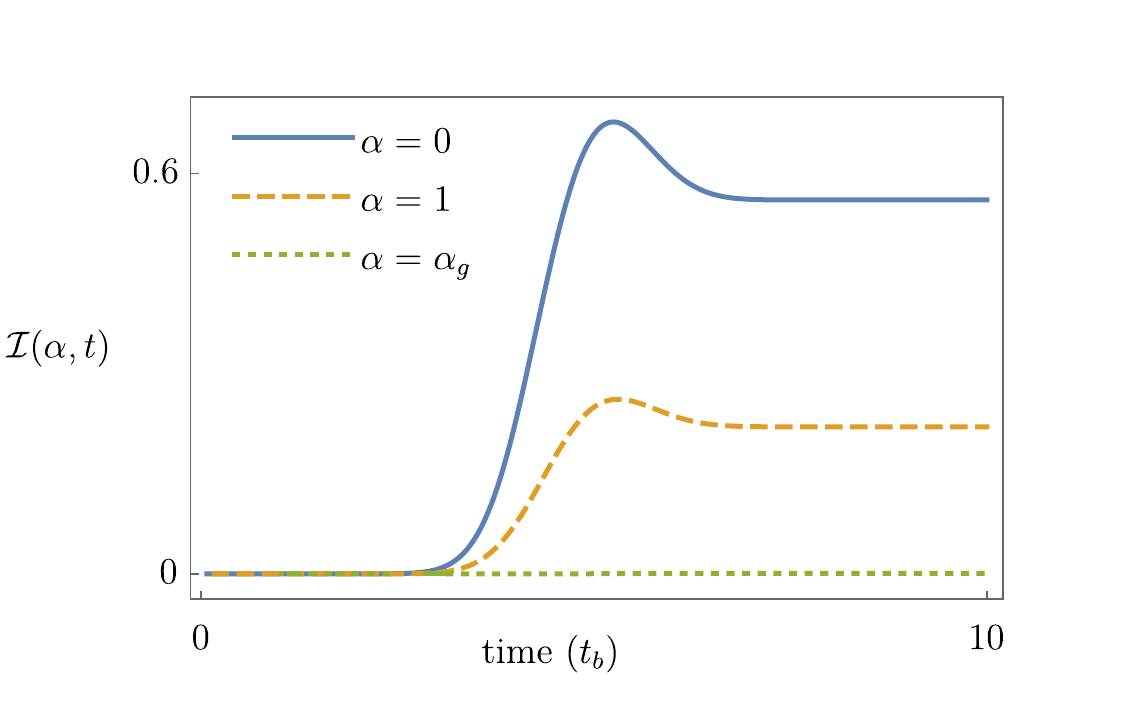}
\caption{The mutual information ${\cal I}(\alpha,t)$ is plotted with time in units of $t_b=w_c/\nu$ assuming an initial state $\ket{0,0}$. The beam waist is $w_c=20~\mu$m, $\omega_m$ is chosen in the microwave regime (energy $\sim10~\mu$eV) and $\nu = 10^{-3}c$ where $c$ is the speed of light. .}\label{mi_gaus}
\end{center}
\end{minipage}
\end{figure}

In the case of uniform motion the Gaussian coupling envelope incurs a relatively smooth switch-on. For a beam waist $w_c = 20~\mu$m, with $h$ substantially larger, so that the dipole starts well outside of the cavity, and for a dipole with microwave frequency $\omega_m\sim$GHz, the gross dipolar speed must be around $\nu = 10^{-3}c$ in order that the interaction time $\tau\sim w_c/\nu$ is comparable to the cycle time $1/\omega_m$. The velocity $10^{-3}c$ is not yet relativistic, but significantly larger than the velocities found in typical atomic beam experiments, which are around three orders of magnitude smaller. In order to achieve $w_c\omega_m/\nu  \sim 1$ with smaller $\nu$ either the cavity beam waist must be further reduced, or slower dipolar oscillations must be considered. However $w_c\omega_m/\nu  \sim 1$ is achieved, significant differences occur in predictions associated with different gauges within this regime, as shown in Fig. \ref{mi_gaus}.

\section{Ground state of the interacting Hamiltonian, and the ground state photon number and mutual information}\label{Gph}

A naive example of a time-dependent interaction comprises instantaneous interaction switch-on/off described by the function $\mu(t) =u(t-t_0)-u(t-(t_0+\tau))$ where $u$ denotes the unit-step function. For final times $t>t_0+\tau$ the evolution of the system is composed of sequential evolutions as $U(0,t)=U_0(0,t_0)U^\alpha(t_0,t_0+\tau)U_0(t_0+\tau,t)$ where $U^\alpha(t',t)=e^{-i(t-t')H^\alpha}$ and $U_0(t',t)=e^{-i(t-t')H_0}$. However, the free (uncoupled) evolution $U_0$ does not alter either the oscillator populations nor the final light-matter correlations. To find these observables one can set $t_0=0$ and $t=\tau$ without loss of generality, which is equivalent to considering the full interacting system with a constant interaction $\mu(t)=1$. In this case it is more physically relevant to consider an initial eigenstate of the full Hamiltonian $H^\alpha$ rather than the free part $H_0$.
\begin{figure}[H]
\begin{minipage}{\columnwidth}
\begin{center}
\hspace*{-0.2cm}\includegraphics[scale=0.43]{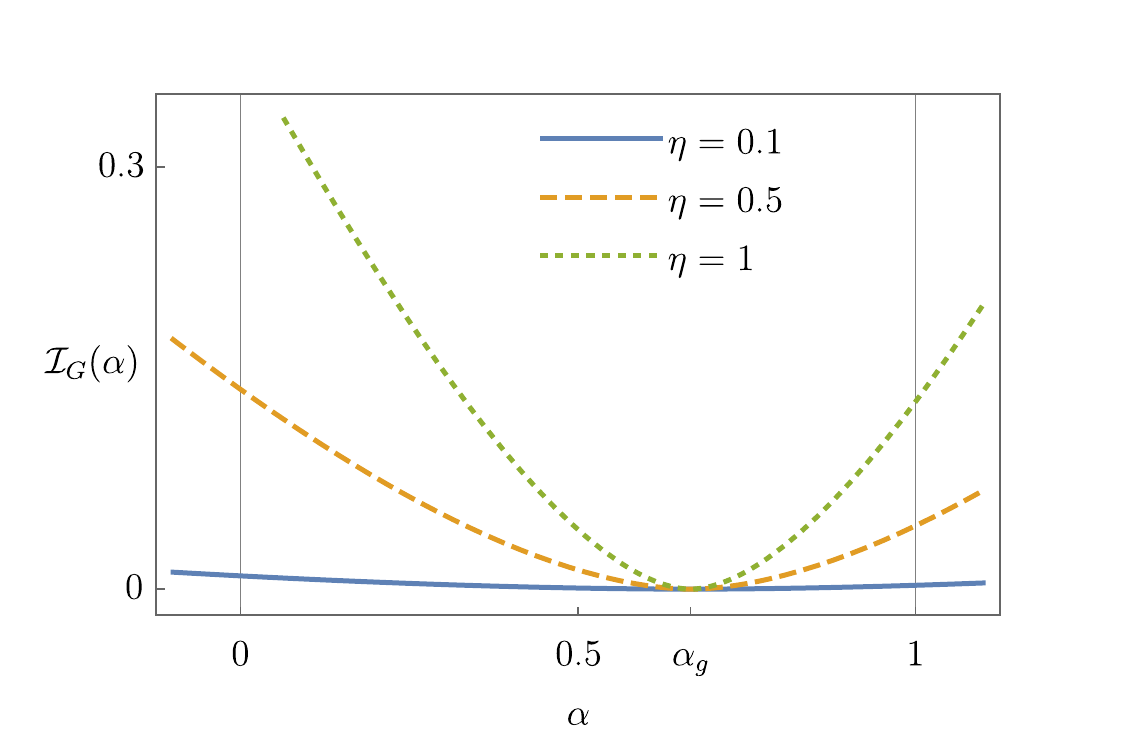}
\caption{${\cal I}_G(\alpha)$ is plotted as a function of $\alpha$ with $\delta=\omega/\omega_m=1/2$, for three values of the dimensionless coupling parameter $\eta=e/(\omega\sqrt{mv})$. The strength of the $\alpha$-dependence increases with increasing $\eta$. For all $\eta$ the mutual information ${\cal I}_G(\alpha)$ is symmetric about the minimum value of zero occurring at $\alpha_g=1/(1+\delta)$ for which the ground state $\ket{G}$ is in fact separable (Appendix \ref{Gph}). At resonance $\delta=1$ we have $\alpha_g=1/2$, implying ${\cal I}_G(0)={\cal I}_G(1)$. Off-resonant values of $\delta$ determine the shift of the minimum $\alpha_g$ relative to the resonant value; $\alpha_g$ is shifted towards $\alpha=1$ for $\delta<1$, and towards $\alpha=0$ for $\delta>1$.}\label{mi}
\end{center}
\end{minipage}
\end{figure}
\begin{figure}[H]
\begin{minipage}{\columnwidth}
\begin{center}
\hspace*{-0.2cm}\includegraphics[scale=0.44]{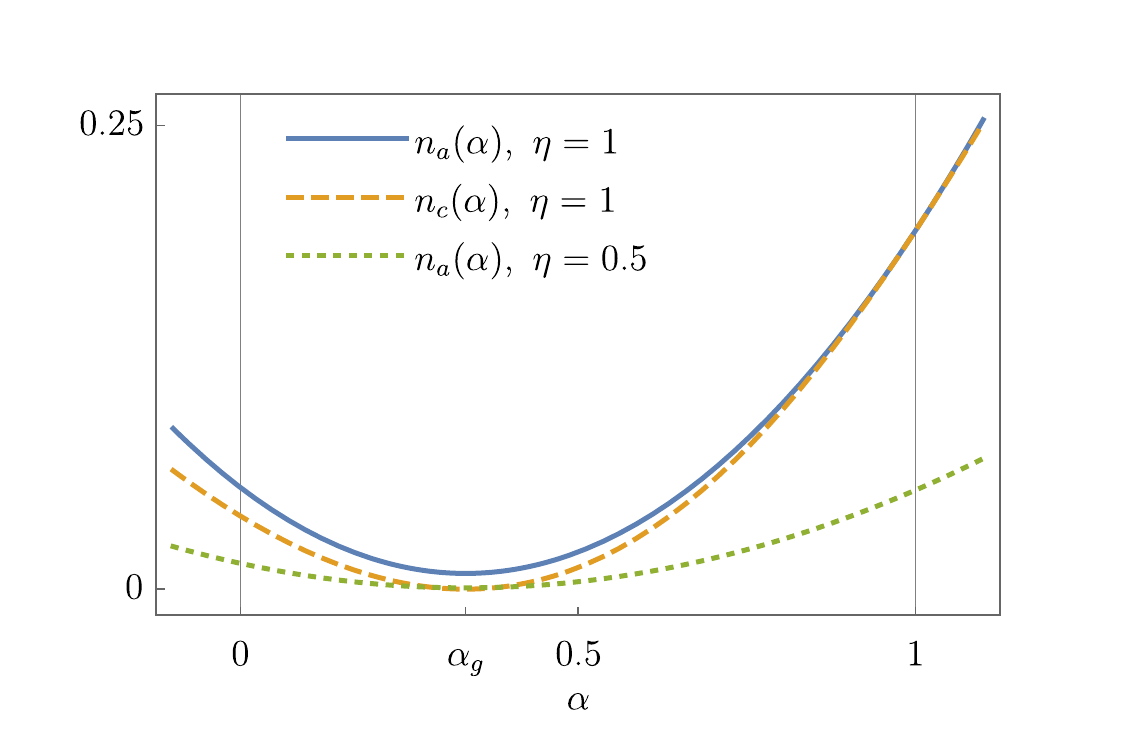}
\caption{The ground state photon number averages $n_a(\alpha)$ and $n_c(\alpha)$ are plotted as functions of $\alpha$ with $\delta=\omega/\omega_m=2$.  The strength of the dependence on $\alpha$ increases with increasing $\eta$, as does the difference between the two photon numbers $n_a$ and $n_c$. For sufficiently weak coupling $\eta\leq 0.1$, $n_a$ and $n_c$ are indistinguishable within the resolution of the plot. Both $n_a$ and $n_c$ are minimum at $\alpha=\alpha_g$. For all couplings $n_c$ is identically zero at $\alpha_g$ while $n_a$ becomes non-zero for stronger coupling. For $\alpha\to 1$, $n_a(\alpha) \to n_c(\alpha)$, because the self-energy term $e^2(1-\alpha)^2{\bf A}^2/2m$ vanishes identically in the Poincar\'e gauge $\alpha=1$.}\label{ph}
\end{center}
\end{minipage}
\end{figure}
Of considerable interest are light-matter correlations in the ground state $\ket{G}$ of the full Hamiltonian $H^\alpha$. These are quantified by the mutual information ${\cal I}_G(\alpha) = S(\rho_m^\alpha)+S(\rho_l^\alpha)$ where $S(\rho) = -{\rm tr}\rho \ln \rho$ and the reduced material and cavity states are defined by $\rho_m^\alpha = {\rm tr}_l^\alpha \ket{G}\bra{G}$ and $\rho_l^\alpha = {\rm tr}_m^\alpha \ket{G}\bra{G}$ respectively. The mutual information ${\cal I}_G(\alpha)$ is found to be
\begin{align}
{\cal I}_G(\alpha) = (\mu_\alpha+1) \ln \left({\mu_\alpha +1 \over 2}\right) -(\mu_\alpha-1) \ln \left({\mu_\alpha -1 \over 2}\right)
\end{align}
where
\begin{align}
\mu_\alpha = \sqrt{1+\left({\omega \over \omega_g}\right)^2 {e^2\over mv \omega\, \omega_m}(\alpha-\alpha_g)^2}.
\end{align}
It is symmetric about the point $\alpha=\alpha_g$ where it takes its minimum value of zero.

We also consider the average number of $\alpha$-gauge photons $n_a(\alpha)=\langle a^\dagger a \rangle_G$ in the ground state. A straightforward calculation yields
\begin{align}\label{av1}
n_a(\alpha)= {1\over 4\omega}\left[\omega_g+{e^2(\alpha-\alpha_g)^2\over mv\omega_{m,g}}+{\omega^2 \over \omega_g}\right]-{1\over 2}
\end{align}
where $\omega_g\equiv \omega_{\alpha_g}$ and
\begin{align}
&\alpha_g ={\omega_m \over \omega_m+\omega},\qquad \omega_{m,g}^2 = \omega_m^2 + {e^2\over mv}\alpha_g^2.
\end{align}
If one allows the definition of photon number to depend on material parameters $e$ and $m$ then the self-energy term
\begin{align}
e^2(1-\alpha)^2{\bf A}^2/2m = \eta^2\omega[(1-\alpha)^2(a^\dagger+a)^2]
\end{align}
can be absorbed into a redefinition of the local cavity energy as
\begin{align}
{\tilde H}_l^\alpha&= H_l^\alpha + e^2(1-\alpha)^2{\bf A}^2/2m = {v\over 2}({\bf \Pi} +\omega_\alpha^2{\bf A}^2) =\omega_\alpha\left(c^\dagger c +{1\over 2}\right)
\end{align}
where
\begin{align}
\omega_\alpha^2 = \omega^2 +{e^2\over mv}(1-\alpha)^2.
\end{align}
The operators $c,~c^\dagger$ are related to $a,\,a^\dagger$ by a local Bogoliubov transformation in ${\cal H}_l^\alpha$. The average number of $\alpha$-gauge renormalised ground state photons $n_c(\alpha)=\langle c^\dagger c \rangle_G$ is
\begin{align}\label{rennum}
n_c(\alpha) = {1\over 4\omega_\alpha}\left[\omega_g+{e^2(\alpha-\alpha_g)^2\over mv\omega_{m,g}}+{\omega_\alpha^2 \over \omega_g}\right]-{1\over 2}.
\end{align}
Unlike the average in Eq.~(\ref{av1}) this average reaches a minimum of zero for $\alpha=\alpha_g$. This can be understood by noting that for this choice of $\alpha$ the Hamiltonian can be written in number-conserving form as
\begin{align}\label{nc}
H^g=\omega_{m,g}\left(d^\dagger d +{1\over 2}\right)+\omega_g\left( c^\dagger c +{1\over 2}\right)+ ie\sqrt{\omega \omega_m \over mv} {1\over \omega_m+\omega}(d^\dagger c -d c^\dagger),
\end{align}
where the renormalised material modes $d$ are such that
\begin{align}
{{\bf p}^2\over 2m} +{m\omega_{m,g}^2 \over 2}{\bf r}^2 = \omega_{m,g}\left(d^\dagger d +{1\over 2}\right).
\end{align}
The renormalised material modes $d$ are connected to the bare material modes $b$ via a local Bogoliubov transformation in ${\cal H}_m^g$. The ground state $\ket{G}$ of the Hamiltonian is the vacuum $\ket{0^d ,0^c}$ annihilated by the operators $d$ and $c$. Thus, $n_c(\alpha_g)=0$. It is important to note that unlike  a full diagonalisation of the Hamiltonian, the partially diagonal form Eq.~(\ref{nc}) does not obscure the divisibility of the overall system into ``light" and ``matter" subsystems. After a full diagonalisation the Hamiltonian can be written as the sum of two harmonic oscillator energies, but it is not possible to distinguish these harmonic oscillators such that one can be called ``light" and the other ``matter" in any meaningful way. This is because a completely diagonalising transformation is necessarily non-local with respect to the light-matter Hilbert space bipartition of any gauge. On the other hand the number-conserving form Eq.~(\ref{nc}) can be achieved by simply choosing a particular gauge and then performing nothing but {\em local} operations within that gauge.

Figure~\ref{mi} shows significant variations in the mutual information ${\cal I}_G(\alpha)$, which become increasingly pronounced for larger dimensionless coupling-strengths $\eta$. Similarly Fig.~\ref{ph} plots $n_a(\alpha)$ and $n_c(\alpha)$ as a functions of $\alpha$, showing that both non-renormalised and renormalised photon numbers vary significantly with $\alpha$.

\end{document}